\documentclass[aps,preprint,showpacs,superscriptaddress,groupedaddress]{revtex4}  
\usepackage{graphicx}  
\usepackage{float}
\usepackage{dcolumn}   
\usepackage{bm}        
\usepackage{amssymb}   
\usepackage{slashed}   
\usepackage{amsmath}   
\usepackage{simplewick} 
\usepackage{verbatim}  
\usepackage{color} 
\usepackage{appendix} 
\usepackage{subfig}

\usepackage[bookmarks=true,colorlinks=true,linkcolor=blue,unicode=true]{hyperref}

\begin{document}

\widetext

\title{Generation of magnetic skyrmions through pinning effect}

\author{Ji-Chong Yang} \affiliation{Department of Physics  \&  State Key Laboratory of Surface Physics,  Fudan University, Shanghai 200433, China}

\author{Qing-Qing Mao} \affiliation{Department of Physics  \&  State Key Laboratory of Surface Physics,  Fudan University, Shanghai 200433, China}

\author{Yu Shi} \affiliation{Department of Physics  \&  State Key Laboratory of Surface Physics,  Fudan University, Shanghai 200433, China}

\affiliation{Collaborative Innovation Center of Advanced Microstructures, Fudan University, \\Shanghai 200433, China}
\vskip 0.25cm

\date{\today}

\begin{abstract}
Based on analytical estimation and lattice simulation, a proposal is made that  magnetic skyrmions  can be generated through the pinning effect in 2D chiral magnetic materials,  in absence of an  external magnetic field or magnetic anisotropy. In our simulation, stable magnetic skyrmions can be  generated in the pinning areas.  The properties of the skyrmions are studied for various values of  ferromagnetic exchange  strength   and the Dzyaloshinskii-Moriya interaction strength.
\end{abstract}

\pacs{75.70.Kw, 66.30.Lw, 75.10.Hk, 75.40.Mg}
\maketitle

\section{\label{sec:1}Introduction}

The topologically protected structure called skyrmion can be formed in a chiral magnet~\cite{nagaosa,Fert}. Magnetic skyrmions have been discovered in the bulk MnSi by using neutron scattering~\cite{Muhlbauer}, and have also been observed by using Lorentz transmission electron microscopy~\cite{Yu1} and by using spin-resolved scanning tunnelling microscopy~(STM)~\cite{2DSkyrmion2}. They can be created in magnetic materials with  Dzyaloshinskii-Moriya~(DM) interactions~\cite{DMI}. They can be driven by  spin current with critical current density lower than that for magnetic domain walls~\cite{ultralow}, and are thus promising as  future information carriers in magnetic information storage and processing devices.

Therefore it is interesting to find efficient methods of creation and manipulation of magnetic skyrmions. In the presence of a magnetic field, single skyrmions can be created and deleted by using local spin-polarized STM~\cite{Romming}. A large number of magnetic skyrmons were created with the aid of a special geometrical constriction in presence of interfacial DM interaction~\cite{bubble,thetarho}. In the absence of a magnetic field,  skyrmions can be generated with the aid of a circulating current~\cite{Tchoe}, or  magnetic anisotropy~\cite{anisotropy,bilayer1},  or  perpendicular anisotropy energy~\cite{anisotropyenergy}, or DC current together with inhomogeneous magnetization but without DM interactions~\cite{dccurrent}.

Pinning effect refers to the inhomogeneities of the ferromagnetic exchange coupling, the DM interaction and the magnetic anisotropy, which may be caused by defects and impurities~\cite{pin,pin2}. In this paper,  we propose a novel method to generate magnetic skyrmions by exploiting the inhomogeneities  of  the ferromagnetic coupling and the DMI strength.

The rest of the paper is organized as the following. In Sec.~\ref{sec:2}, we briefly review the 2D magnetic skyrmions and introduce our basic idea. In Sec.~\ref{sec:4}, we report a lattice simulation and study the properties of the skyrmions generated in the simulation. A summary is made in  Sec.~\ref{sec:5} .

\section{\label{sec:2}Basic Idea}

In consideration of that  the skyrmions can be generated  by using  magnetic anisotropy, here we only consider the case without magnetic anisotropy. In this case, in the presence of an  external magnetic field,  the Hamiltonian  can be written as~\cite{nagaosa,FreeEnergy}
\begin{equation}
\begin{split}
&{\mathcal H}_{\rm tot}({\bf r})=\frac{J({\bf r})}{2}\left(\nabla {\bf n}\right)^2+D({\bf r}) {\bf n}\cdot (\nabla \times {\bf n})-{\bf B}(\bf r)\cdot {\bf n},\\
\end{split}
\label{eq.2.1}
\end{equation}
where ${\bf n}$ is the orientation of the  magnetic moment, $J({\bf r})$ is the  ferromagnetic exchange strength, $D({\bf r})$ is the strength of DM interaction, ${\bf B}=B\mathbf{e}_z$ is the external magnetic field along ${\bf e}_z$ direction. We use dimensionless parameters.

The local magnetic moment of a skyrmion  or anti-skyrmion can be parameterized as~\cite{nagaosa}.
\begin{equation}
\begin{split}
&{\bf n}(x,y)=\left(\cos(\gamma +m \phi)\sin(\theta(\rho)),\sin(\gamma +m \phi)\sin(\theta(\rho)),g \cos (\theta (\rho))\right),\\
\end{split}
\label{eq.2.2}
\end{equation}
where $\rho$ and $\phi$ are polar coordinates of the 2D position  vector $(x, y)$, with the origin at the center of the skyrmion, $m=\pm 1$ is the vorticity of the skyrmion  or anti-skyrmion, $g=\pm 1$  is the polarization, $\gamma$ is helicity and   distinguishes  between N$\rm \acute{e}$el-type and Bloch-type skyrmions, $\theta (\rho)$ is a function  describing  the shape of a skyrmion,  with $\theta(0)=\pi$ and $\theta(\infty)=0$. The skyrmion charge is defined as~\cite{nagaosa,Tchoe}
\begin{equation}
\begin{split}
&Q=\frac{1}{4\pi}\int dxdy {\bf n}\cdot \left(\frac{\partial {\bf n}}{\partial x}\times \frac{\partial {\bf n}}{\partial y}\right)=-mg.\\
\end{split}
\label{eq.2.3}
\end{equation}
The anti-skyrmions usually result from anisotropic DM interaction~\cite{antiskyrmion}, but here  we only consider isotropic DM interaction, hence  we  only consider  $m=1$.

\begin{figure}
\centering
\subfloat[$g=1$,$m=1$,$\gamma=0$]{\includegraphics[width=0.25\textwidth]{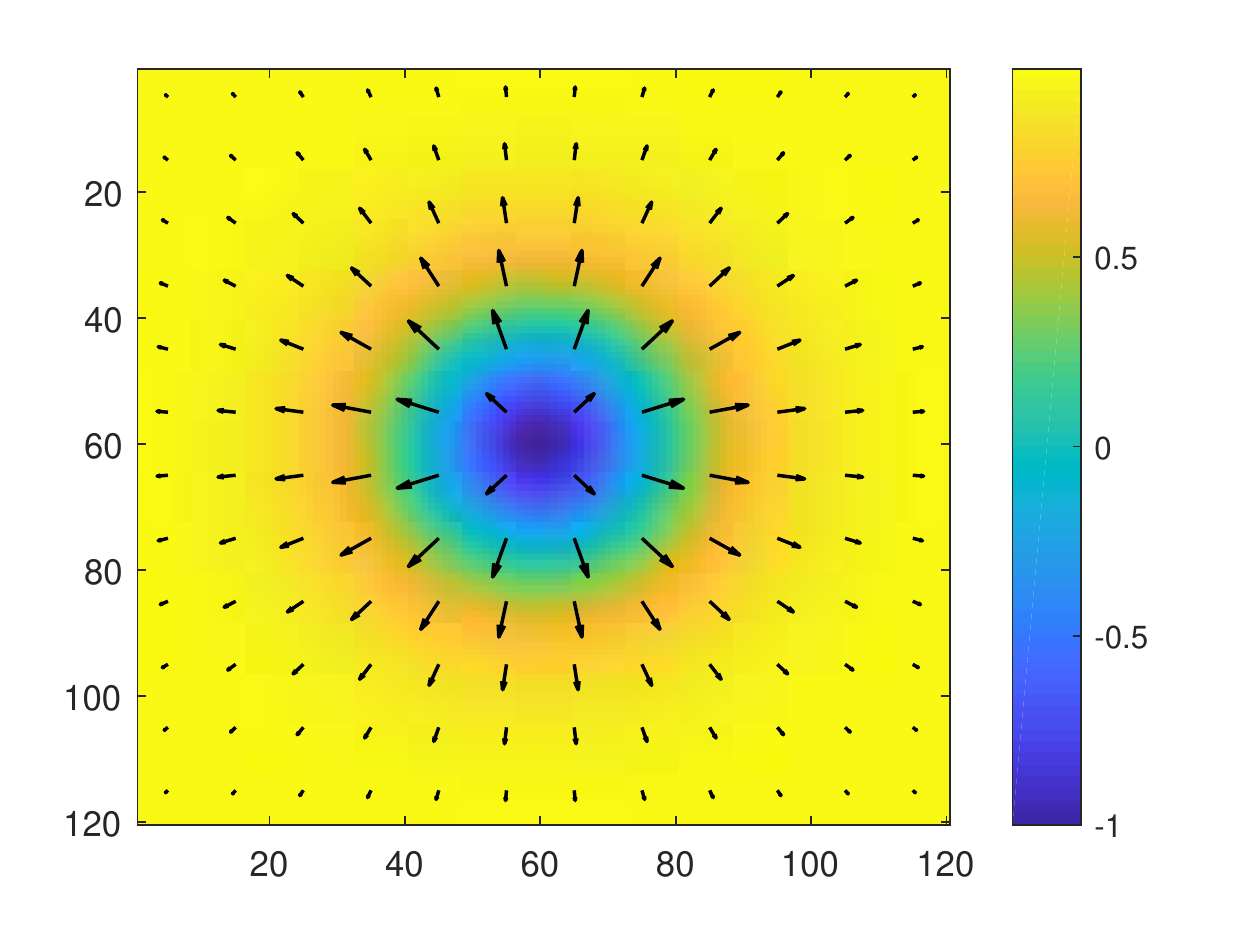}}\hfill
\subfloat[$g=1$,$m=1$,$\gamma=\frac{\pi}{2}$]{\includegraphics[width=0.25\textwidth]{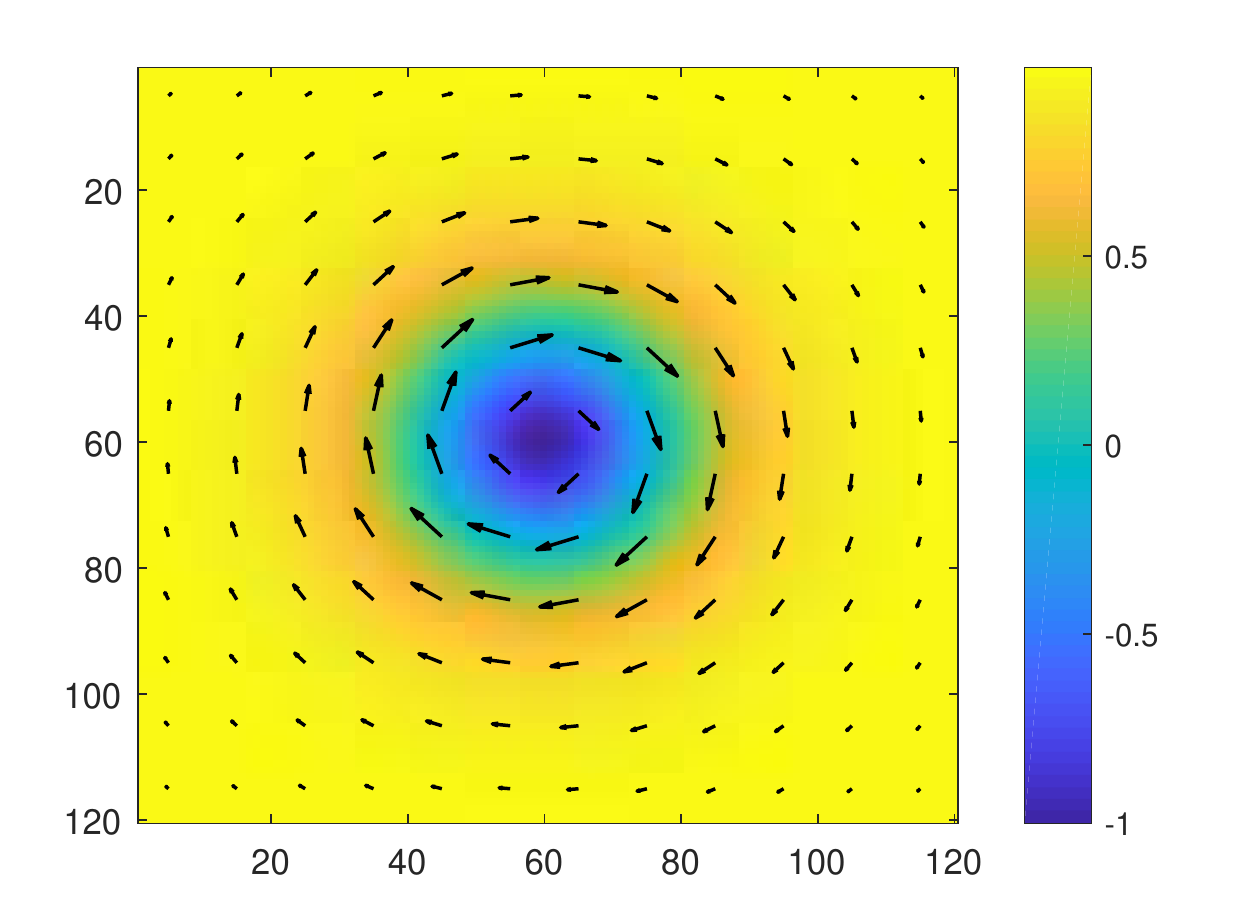}}\hfill
\subfloat[$g=-1$,$m=1$,$\gamma=0$]{\includegraphics[width=0.25\textwidth]{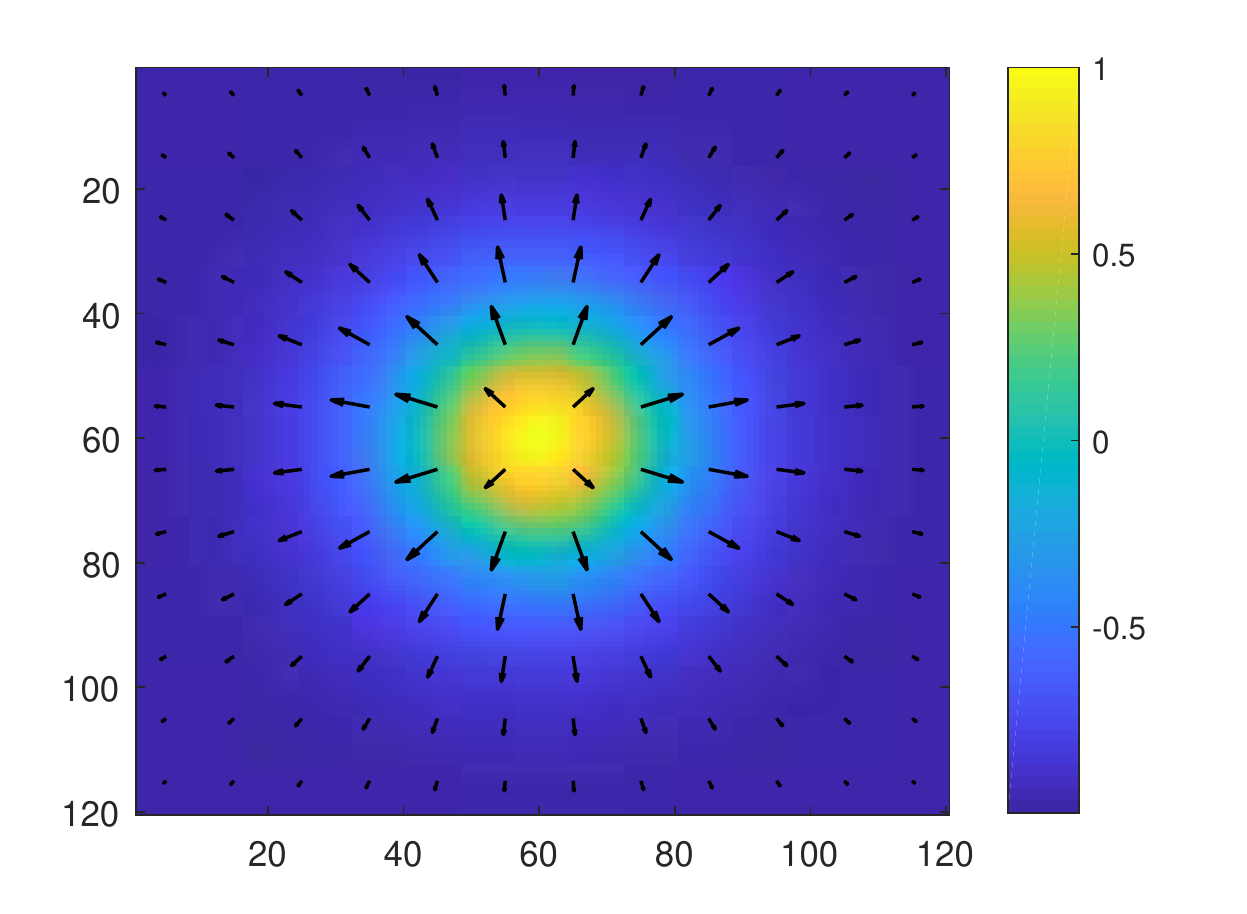}}\hfill
\subfloat[$g=-1$,$m=1$,$\gamma=\frac{\pi}{2}$]{\includegraphics[width=0.25\textwidth]{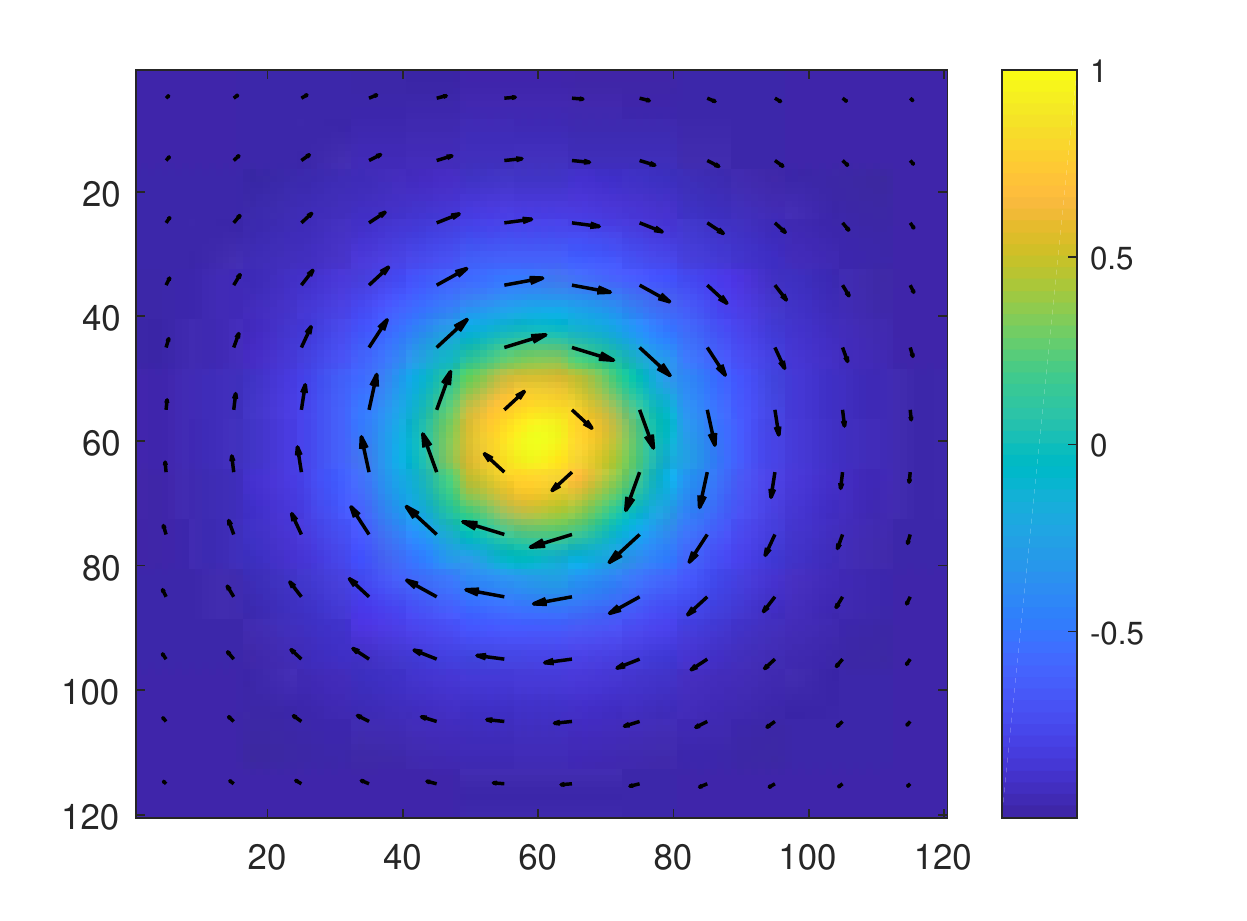}}
\caption{Skyrmions characterized by ${\bf n}(\bf r)$ parameterized as in Eqs.~(\ref{eq.2.2}) and (\ref{eq.2.5}),  with various values of $g$ and $\gamma$. The heat map represents the magnitude of $n_z$ and the arrow  represents  $(n_x,n_y)$. }
\label{fig:examples}
\end{figure}

It has been found that for a skyrmion,  $\theta(\rho)$ can be approximated as~\cite{arctan}
\begin{equation}
\begin{split}
&\theta(\rho)\approx 4\tan^{-1}(\exp(-a\rho)).\\
\end{split}
\label{eq.2.5}
\end{equation}
Obtained from this expression with $a=0.05$,  some examples of  ${\bf n}({\bf r})$ are shown in Fig.~\ref{fig:examples}.

Two kinds of pinning were considered  previously. In Ref.~\cite{pin}, both $J$ and $D$ are inhomogeneous, while $D/J$ is kept constant. In Ref.~\cite{pin2}, $D$ is constant while $J$ is inhomogeneous. For simplicity, we only consider the latter case in this section. The former case will be studied in a lattice simulation in next section.

From  Eq.~(\ref{eq.2.1}), we obtain the Euler-Lagrange equation
\begin{equation}
\begin{split}
&\frac{ \sin(\theta)\cos(\theta)}{\rho}-\theta ' - \rho \theta ''-2\frac{\tilde{D}}{J(\rho)}\sin^2(\theta)+\frac{\tilde{B}}{J(\rho)} \rho \sin (\theta)-\rho \frac{J'(\rho)}{J(\rho)}\theta '=0,
\end{split}
\label{eq.2.6}
\end{equation}
where $\tilde{D} \equiv g\sin(\gamma)D$, $\tilde{B}\equiv gB$, $J(\rho)$ is a pinning function, simply assumed to depend only on $\rho$  so that the skyrmion center is at the center of the pinning, which is  rotationally symmetric. When there is no pinning while an external magnetic field is applied,  Eq.~(\ref{eq.2.6}) becomes
\begin{equation}
\begin{split}
&\frac{ \sin(\theta)\cos(\theta)}{\rho}-\theta ' - \rho \theta ''-2\frac{\tilde{D}}{J}\sin^2(\theta)+\frac{\tilde{B}}{J} \rho \sin (\theta)=0.\\
\end{split}
\label{eq.2.7}
\end{equation}
It is known that the skyrmions can be generated  in this case.

We note that when pinning is present,  $J'$ term in Eq.~(\ref{eq.2.6})   plays a role similar to that of   $\tilde{B}$ term. Using the ansatz $\theta(\rho)$ in Eq.~(\ref{eq.2.5}) for $b  \sin(\theta(\rho))=J'(\rho)\theta'(\rho)$, where $b$ is some value of $\tilde{B}$ ,  we find that if  \begin{equation}
\begin{split}
&J(\rho)=J_0+b\frac{a\rho+\log \left(e^{-2a\rho}+1\right)}{a^2},
\end{split}
\label{eq.2.8}
\end{equation}
where $a$ as given in Eq.~(\ref{eq.2.5}), $J_0$  is an  undetermined coefficient, then Eq.~(\ref{eq.2.6}) without  $\tilde{B} $ term  becomes
\begin{equation}
\begin{split}
&\frac{ \sin(\theta)\cos(\theta)}{\rho}-\theta ' - \rho \theta ''-2\frac{\tilde{D}}{J(\rho)}\sin^2(\theta)+\frac{b}{J(\rho)} \rho \sin (\theta)=0,\\
\end{split}
\label{eq.2.9}
\end{equation}
which is very close to Eq.~(\ref{eq.2.7}) except that the $J$ is now position dependent.

\begin{figure}
\includegraphics[width=0.7\textwidth]{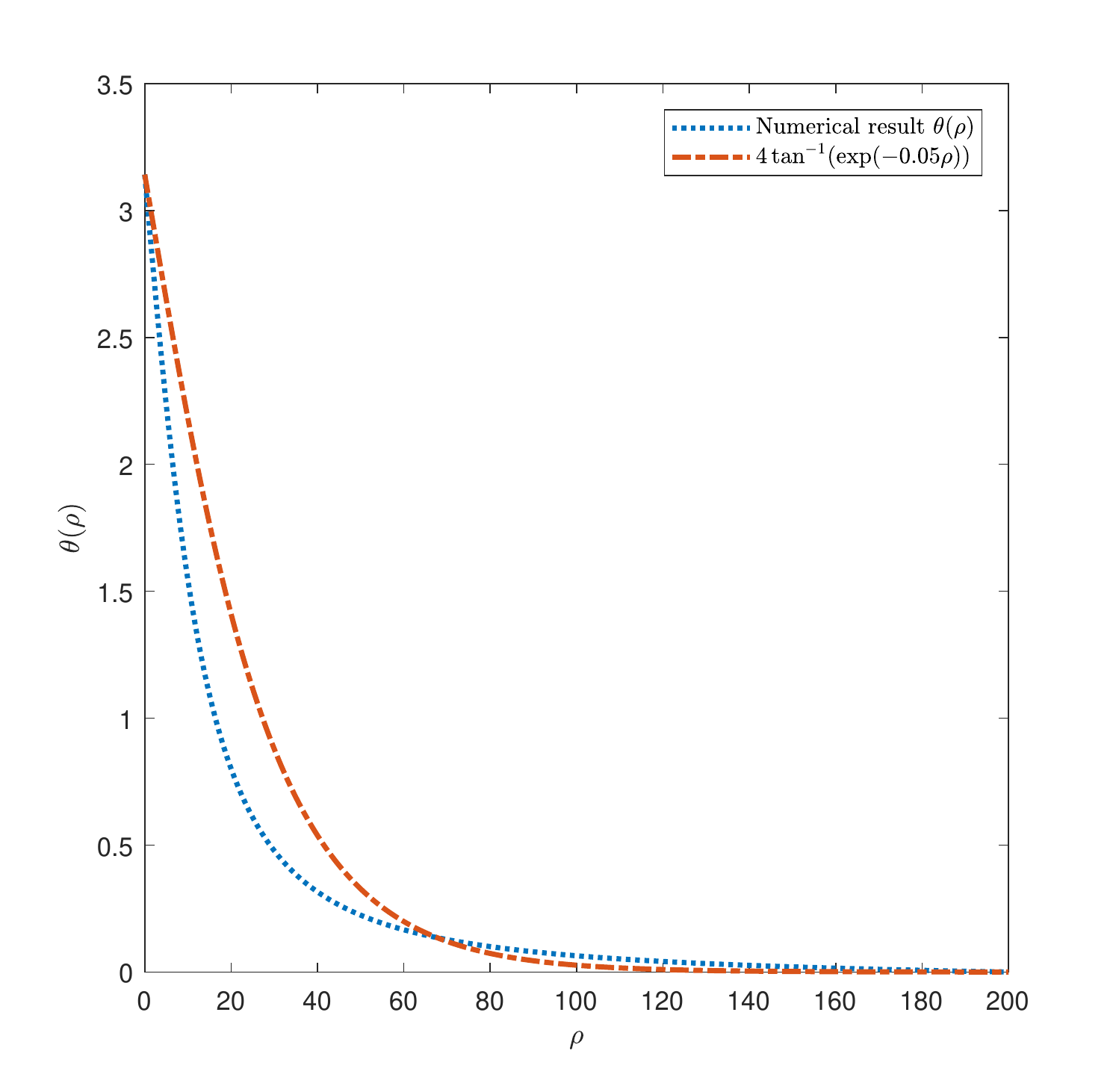}
\caption{The dotted line represents the numerical result  of Euler-Lagrange Equation (\ref{eq.2.6}) with $\tilde{B}=0$. The dotted-dashed line represents the skyrmion ansatz  (\ref{eq.2.5}).}
\label{fig:ansatz}
\end{figure}

With $J(\rho)$  in Eq.~(\ref{eq.2.8}) as an ansatz, we solve Eq.~(\ref{eq.2.6}) in the absence of   $\tilde{B} $ term,  using the numerical method in Ref.~\cite{pin}, which proposed a mechanism of pinning skyrmions.   We choose  the parameter values to be $J_0=1$, $\tilde{B}=0$, $\tilde{D}=0.05$, $b=0.005$, $a=0.05$. As shown in Fig.~\ref{fig:ansatz}, the solution is very close to the skyrmion ansatz  (\ref{eq.2.5}), suggesting that it is possible to create  skyrmions by using the pinning effect.

\section{\label{sec:4}Lattice simulation}

The above estimation with an artificial pinning effect suggests the possibility of really creating a skyrmion using pinning effect only. Since the effective parameter $\tilde{D}$  depends not only on  $D$ but also on the parameters of the skyrmions, whether or not a solution of $\theta(\rho)$ can be identified as a skyrmion is technically subtle. Therefore, rather than solving the Euler-Lagrange equation,  we perform a lattice simulation,  which directly provides evidence of skyrmions.

We now consider more realistic pinning effect. As a local structure, the effect of pinning should be  suppressed very quickly in deviating away from the pinning center.  An exponentially decaying  function $J(\rho)$ is assumed in Ref.~\cite{pin2}, while a Gaussian  function $J(\rho)$  is assumed in Ref.~\cite{pin}. We follow Ref.~\cite{pin} to assume $J(\rho)$ to be Gaussian,
\begin{equation}
\begin{split}
&J(\rho)=J_0+J_1e^{-J_2\rho^2},
\end{split}
\label{eq.3.1}
\end{equation}
where $J_0$, $J_1$ and $J_2$ are undetermined coefficients,  with $J_0>0$, $J_1>-J_0$ and $J_2>0$. The radius of the pinning is denoted as $R_p$, and  $R_p \sim 1/\sqrt{J_2}$.

In the simulation, as dimensionless parameters,  $J=1$  is used as the definition of the energy unit,  $J(\infty)=J_0=1$ is assumed for simplicity~\cite{unitJ,alpha3,bilayer}. The dimensionless quantities can be rescaled to physical ones as the following~\cite{Tchoe,unitJ,pin}.  The time unit  $\Delta t = 1$ in the simulation represents  $t=\hbar / J $ in physical time.  The rescaling factor $r$ can be determined from the helical wavelength $\lambda$ and the lattice spacing $a$,  as $r=(D/J)\lambda/(2\pi \sqrt{2}a)$. Then the time is rescaled  as $t'=r^2 t$.
For example, if we adopt the real material such that $\lambda \approx 60 \;{\rm nm}$ and $a\approx 4\;{\AA}$ and with $D/J=0.5$, we find $r\approx 8.44$.
Then, if we adopt the energy unit as $J=3\;{\rm meV}$, then $t'=r^2\hbar/J  \approx 220r^2\;{\rm fs} \approx 0.016\;{\rm ns}$.

The lattice simulation is based on the Landau-Lifshitz-Gilbert~(LLG) equation~\cite{nagaosa,pin,LLG,LLG2}
\begin{equation}
\begin{split}
&\frac{d}{dt} {\bf n}_{\bf r}=-{\bf B}_{\rm eff}({\bf r})\times {\bf n}_{\bf r}-\alpha {\bf n}_{\bf r}\times \frac{d}{dt} {\bf n}_{\bf r},\\
\end{split}
\label{eq.4.1}
\end{equation}
where ${\bf n}_{\bf r}$ is the local magnetic moment at site  ${\bf r}$, $\alpha$ is the Gilbert damping constant, ${\bf B}_{\rm eff}$ is the effective magnetic field,
\begin{equation}
\begin{split}
&{\bf B}_{\rm eff}({\bf r})=-\frac{\delta H}{\delta {\bf n}_{\bf r}},
\end{split}
\label{eq.4.2}
\end{equation}
with the discrete Hamiltonian~\cite{unitJ,discreteH}
\begin{equation}
\begin{split}
&H=\sum _{{\bf r},i=x,y}\left[-J({\bf r}){\bf n}_{{\bf r}+\delta_i}-D({\bf r}) {\bf n}_{{\bf r}+\delta_i}\times {{\bf e}}_i-{\bf B}\right]\cdot {\bf n}_{\bf r},\\
\end{split}
\label{eq.4.3}
\end{equation}
where  $ \delta _{i}$ refers to each neighbour, and  $\delta _{i}= {\bf e}_{i}$ on a square lattice.  So~\cite{pin}
\begin{equation}
\begin{split}
&{\bf B}_{\rm eff}({\bf r})=\sum _{i=x,y}\left[J({\bf r}){\bf n}_{{\bf r}+\delta_i}+J({\bf r}-\delta_i){\bf n}_{{\bf r}-\delta_i}\right]\\
&+\sum _{i=x,y}\left[D({\bf r}){\bf n}_{{\bf r}+\delta_i}\times {{\bf e}}_i-D({\bf r}-\delta _i){\bf n}_{{\bf r}-\delta_i}\times {{\bf e}}_i\right]+{\bf B}({\bf r}).
\end{split}
\label{eq.4.4}
\end{equation}

Unless specified otherwise,  the simulation is run on a $512\times 512$ square lattice with open boundary condition and ${\bf B}=0$, and with   the pinning center  set to be at the point $(256,256)$. In the following, we denote the time step as $\Delta t$, and the time in unit of $\Delta t$ the simulation takes is denoted as $\tau$. The number of step is  $\tau/\Delta t$. The simulation is run on the GPU, which  has a great advantage over CPU on this problem. Because ${\bf n}_{{\bf r}}$'s for different sites at a same instant are independent of each other,  we use GPU to do parallel computing. Within the simulation, the LLG is numerically integrated by using fourth-order Runge-Kutta method.

\subsection{\label{sec:4.1} Skyrmion generation from random  initial configurations}

We first run the simulation for $J_1=3$, $J_2=0.001$ and various values of $D$.  The simulation starts from  randomized ${\bf n}_{\bf r}$'s and stops  when ${\bf n}_{\bf r}$'s become stable. Previously, the Gilbert constant is taken to be $\alpha=0.01$ to $1$~\cite{pin,LLG2,antiskyrmion,arctan,alpha1,alpha2,alpha3,bilayer}. We find that the larger the value of $\alpha$, the more rapid the simulation is completed,  while the smaller the value of $D$, the longer the simulation time. Hence we use $\alpha=0.1$~\cite{LLG2,alpha2} and $\Delta t=0.002$ for $D\geq 0.1$, while  $\alpha=0.2$~\cite{pin,alpha3} and $\Delta t=0.01$ for $D< 0.1$.

\begin{figure}
\subfloat[$D=1$,$\tau=60000\Delta t=120$,$g=1$,$m=1$,$\gamma\approx 1.58$,$R_s\approx 10.5$]{\includegraphics[width=0.33\textwidth]{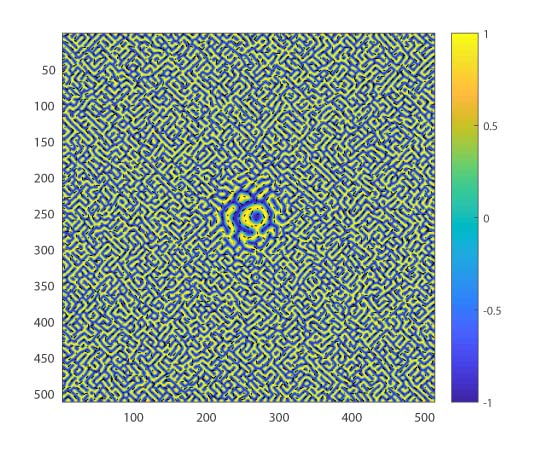}}\hfill
\subfloat[$D=0.8$,$\tau=1000$,$g=1$,$m=1$,$\gamma\approx 1.56$,$R_s\approx 15.4$]{\includegraphics[width=0.33\textwidth]{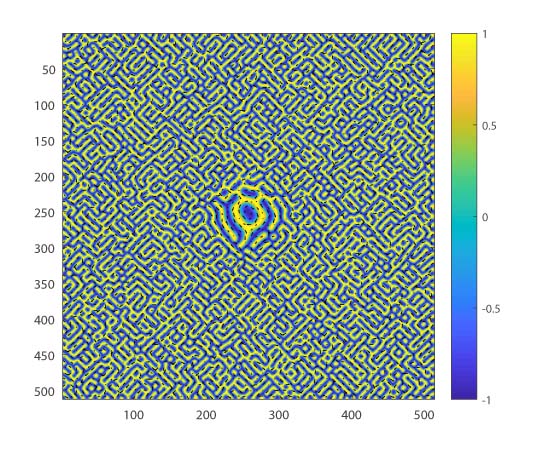}}\hfill
\subfloat[$D=0.5$,$\tau=10000$,$g=1$,$m=1$,$\gamma\approx 1.57$,$R_s\approx 20.3$]{\includegraphics[width=0.33\textwidth]{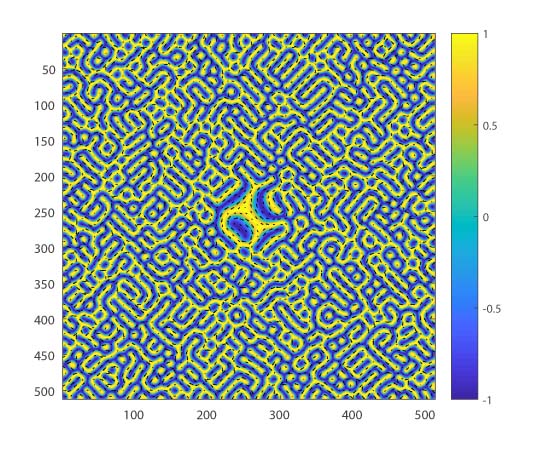}}\vfill
\subfloat[$D=0.3$,$\tau=3000$,$g=1$,$m=1$,$\gamma\approx 1.57$,$R_s\approx 34.4$]{\includegraphics[width=0.33\textwidth]{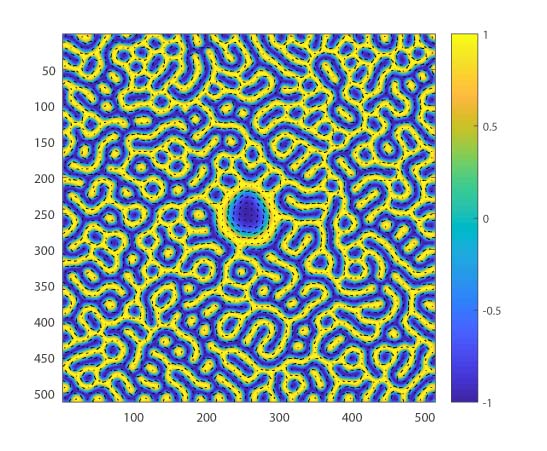}}\hfill
\subfloat[$D=0.2$,$\tau=10000$,$g=-1$,$m=1$,$\gamma\approx 4.72$,$R_s\approx 42.6$]{\includegraphics[width=0.33\textwidth]{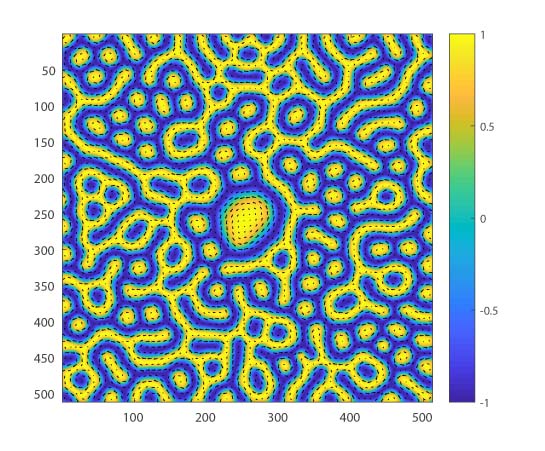}}\hfill
\subfloat[$D=0.1$,$\tau=16000$,$g=-1$,$m=1$,$\gamma\approx 4.71$,$R_s\approx 59.8$]{\includegraphics[width=0.33\textwidth]{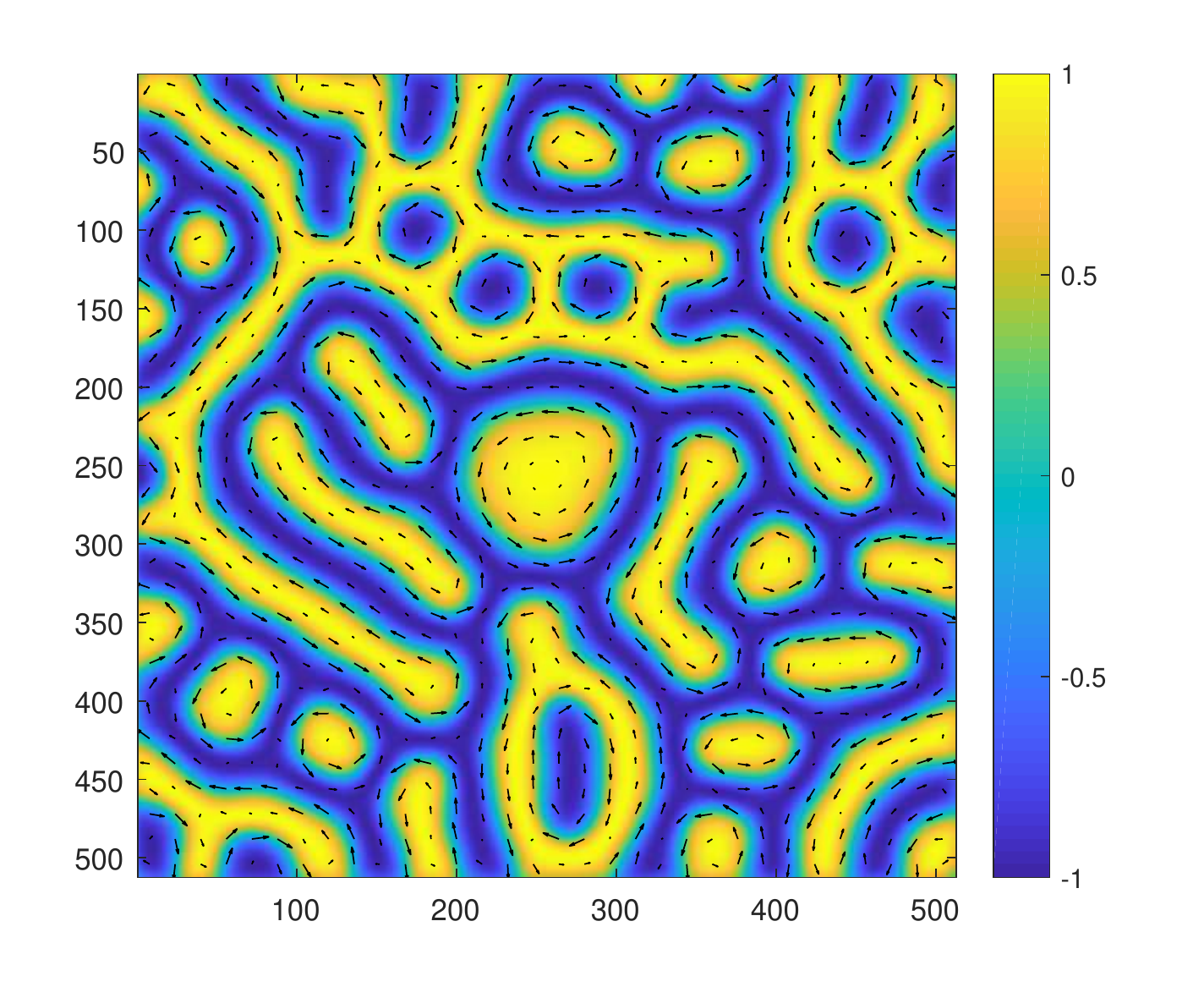}}\vfill
\subfloat[$D=0.08$,$\tau=30000$,$g=-1$,$m=1$,$\gamma\approx 4.71$,$R_s\approx 64.8$]{\includegraphics[width=0.33\textwidth]{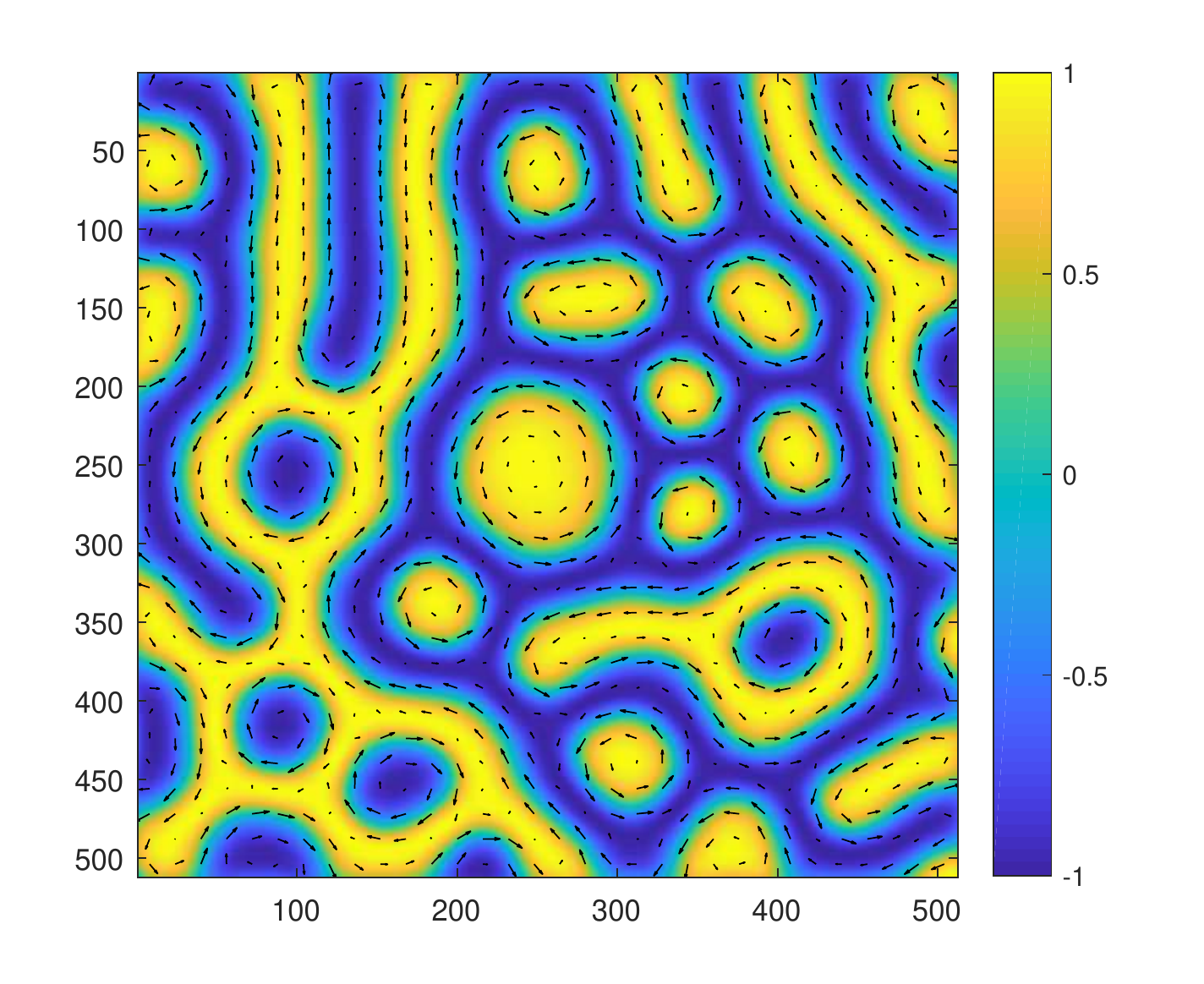}}\hfill
\subfloat[$D=0.05$,$\tau=100000$,$g=1$,$m=1$,$\gamma\approx 1.57$,$R_s\approx 96.7$]{\includegraphics[width=0.33\textwidth]{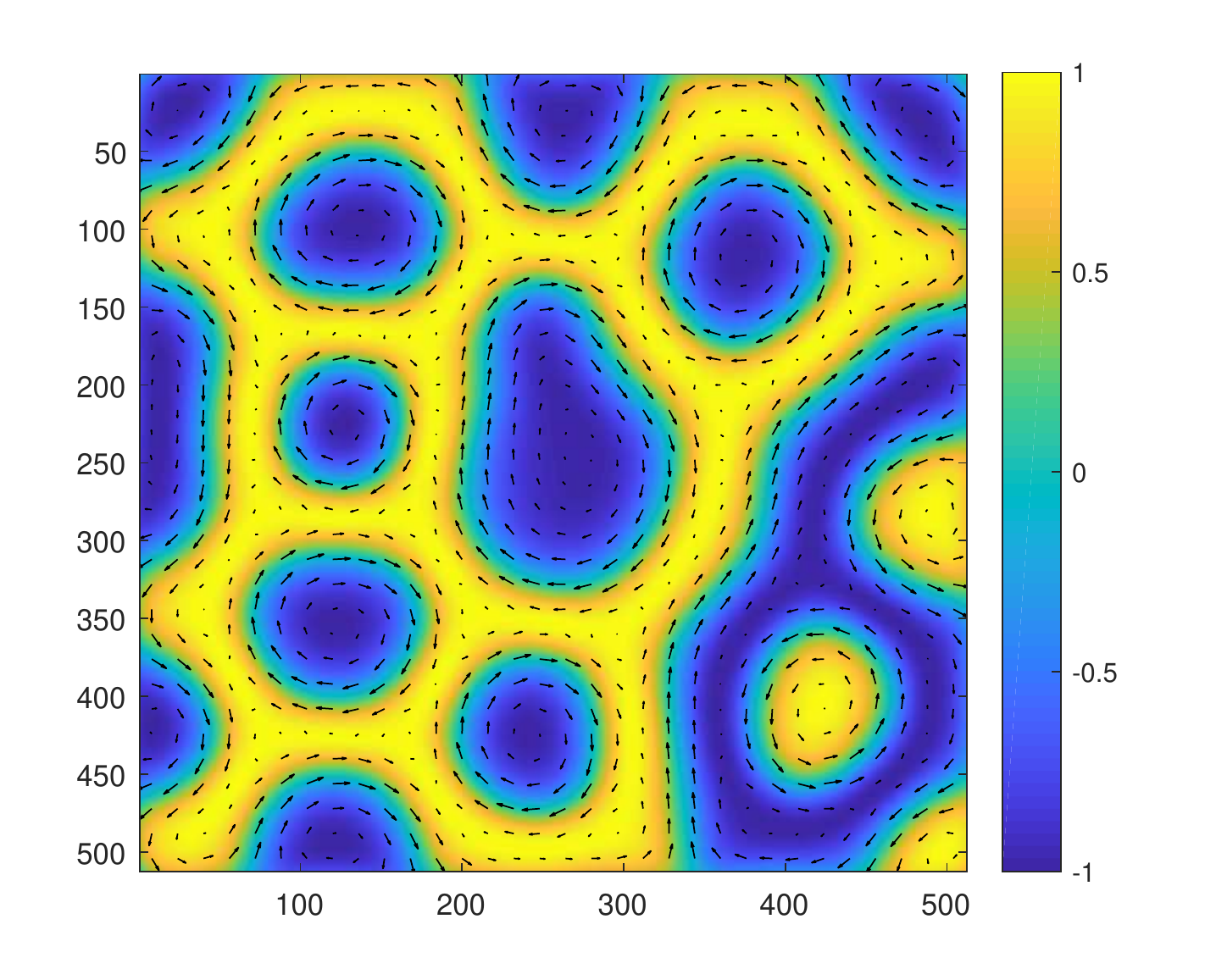}}\hfill
\subfloat[$D=0.03$,$\tau=100000$,$g=-1$,$m=1$,$\gamma\approx 4.71$,$R_s\approx 128.3$]{\includegraphics[width=0.33\textwidth]{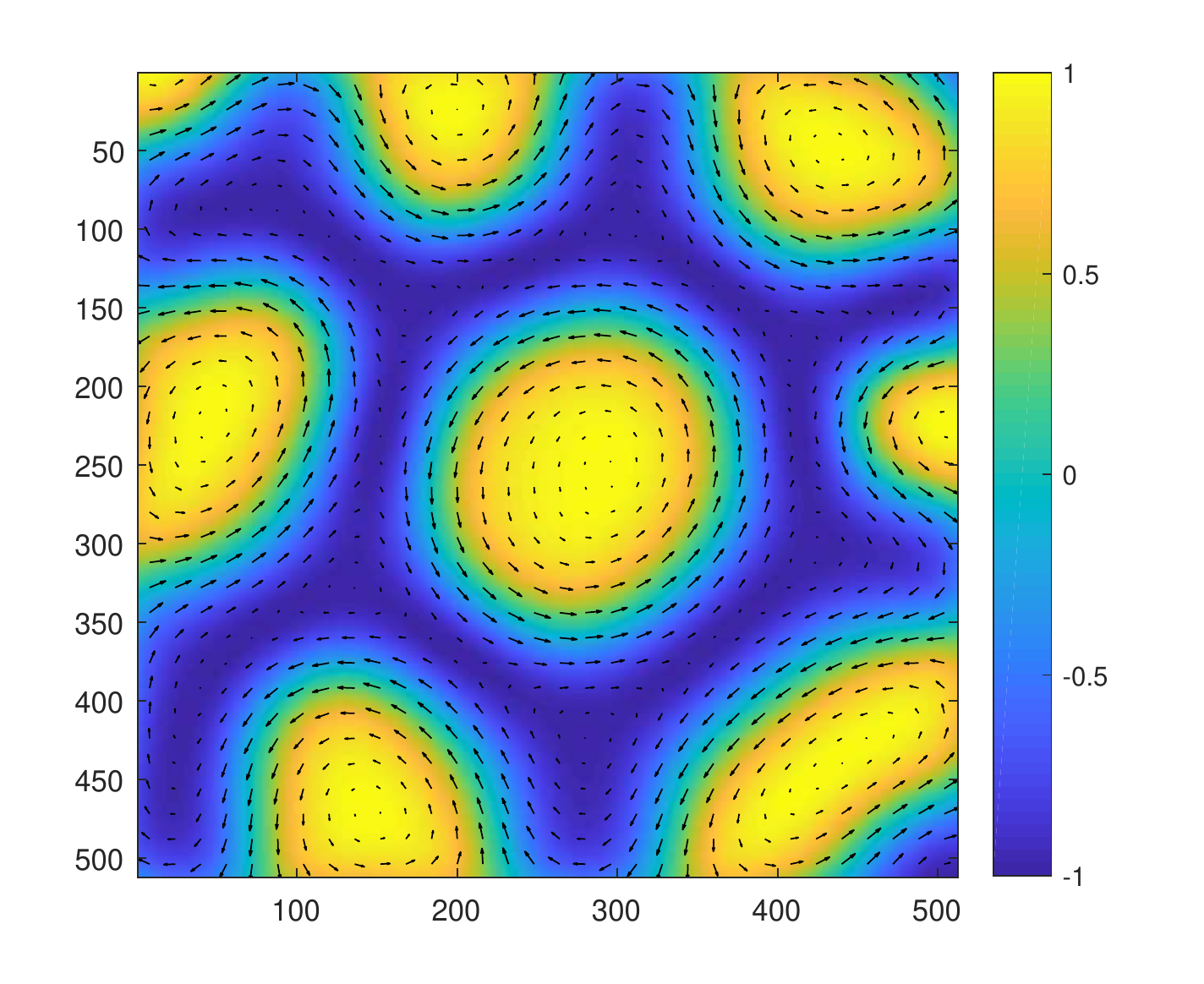}}\vfill
\caption{The stable spin configuration ${\bf n}_r$ in the lattice  simulation, for $J_1=3$ and various values of other parameters. We find that the skyrmions can be generated at the pinning centers, by using the pinning effect only.}
\label{fig:results1}
\end{figure}

The results are shown in Fig.~\ref{fig:results1}. We run the simulation  for $D=1, 0.8, 0.5, 0.2, 0.1, 0.08, 0.05, 0.03, 0.02, 0.01$. For each value of $D$ except the smallest ones $D=0.02$ or $0.01$, a skyrmion can be generated  at the  pinning center. When $D$ is smaller, it takes longer time to generate the skyrmion. Generally speaking, the time needed to generate a skyrmion by using   pinning  is longer than  generation of a skyrmion  by using an  external magnetic field.

We have also studied  the properties of the skyrmions generated in the simulation. Only those skyrmions near    the pinning centers are considered. For each skyrmion considered, $g$ is determined from  $n_z$ at the center. The radius $R_s$ of each skyrmion  can be determined  from the iso-height contour with $n_z=\mp 1$ for $g=\pm 1$ respectively,  because  $n_z$ varies from $\pm 1$ at  the center to $\mp 1$  on  the edge. In the actual  simulation, however, $n_z$ can only approach $\mp 1$. Hence,   $R_s$ is  estimated   from the radius of the iso-height contour with $n_z=\mp 0.9$ for $g=\pm 1$, respectively. $\gamma$ is determined  by using $n_x$ and $n_y$ at the iso-height contour with $n_z=0$.

\begin{figure}
\includegraphics[width=0.7\textwidth]{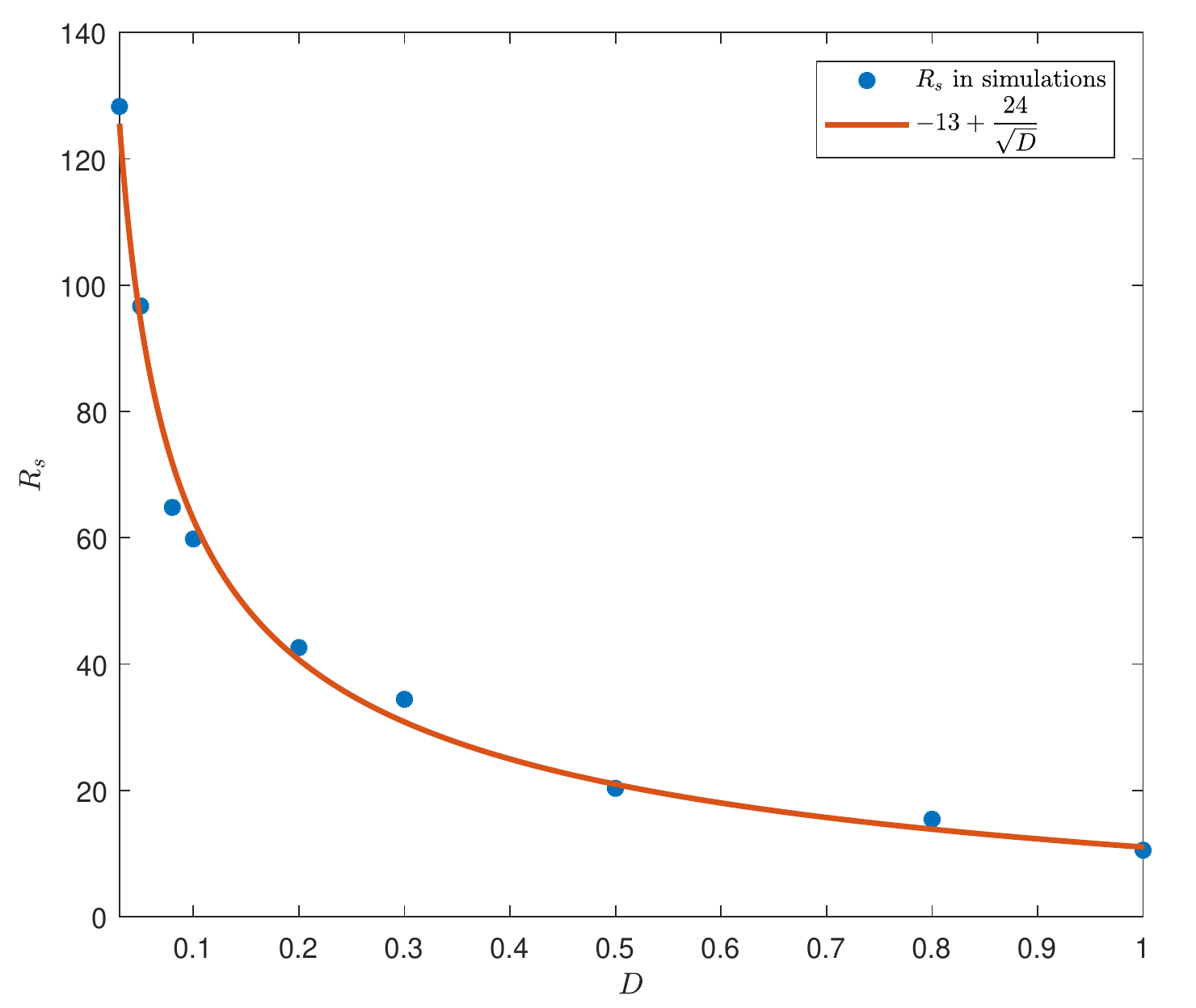}
\caption{The  relation between skyrmion radius $R_s$ and the DM interaction strength $D$.}
\label{fig:rsrelation}
\end{figure}

We first investigate the relation between the skyrmion radius  $R_s$  and the DM interaction strength  $D$. We find that the larger the value of $D$, the smaller  $R_s$. As shown in Fig.~\ref{fig:rsrelation}, the relation between $R_s$  and $D$ is about
\begin{equation}
a+b\frac{1}{\sqrt{D}},
\end{equation}
which explains why   skyrmions are not generated for   $D=0.02$ or  $D=0.01$, as for small values of $D$, the radius of the skyrmion is too large for the lattice size.

All the skyrmions appearing in the simulation are of  $m=1$. Then the skyrmion number $Q=-mg$ is determined from    the sign of $g$. As shown in Fig.~\ref{fig:results1}, skyrmions  with $g= 1$ and $g=-1$ can both be generated. For a skyrmion generated in an external magnetic field, $g$ is determined by the direction of ${\bf B}$.  But for a skyrmion generated by using the pinning effect in absence of  a magnetic field, the sign of $g$ becomes a free choice and  depends on the initial state. This  is verified in  simulations starting with different randomized initial states but with the same values of various parameters. For $D>0$, we  find that   $\gamma \approx \pi/2$   for $g=1$, while  $\gamma \approx  -\pi/2$  for $g=-1$,  hence $\tilde{D}>0$ in both  cases. This can be understood by considering
\begin{equation}
\begin{split}
&H_{\rm DM}=D {\bf n}\cdot \left(\nabla \times {\bf n}\right)=gD\sin \left((m-1)\phi+\gamma\right)\left(\frac{m}{2\rho}\sin (2\theta (\rho))+\theta '(\rho)\right),  \\
\end{split}
\label{eq.4.5}
\end{equation}
which differs from the special case of $g=1$~\cite{nagaosa} in replacing $D$ as $gD$. As a result, the sign of $g$ is a free choice, while the sign of $\gamma$ is determined by $gD$.  Hence the energy is lowest when  $\tilde{D} >0 $ and $\gamma = \pm \pi / 2$ with the sign of $\gamma$ determined by $gD$, that is, $\gamma =   \pi / 2$ when  $gD>0$,   while  $\gamma = -  \pi / 2$ when  $gD<0$.

\begin{figure}
\subfloat[$D=0.03$,$\tau=100000$,$g=1$,$m=1$,$\gamma\approx 1.5727$,$R_s\approx 48.6$]{\includegraphics[width=0.33\textwidth]{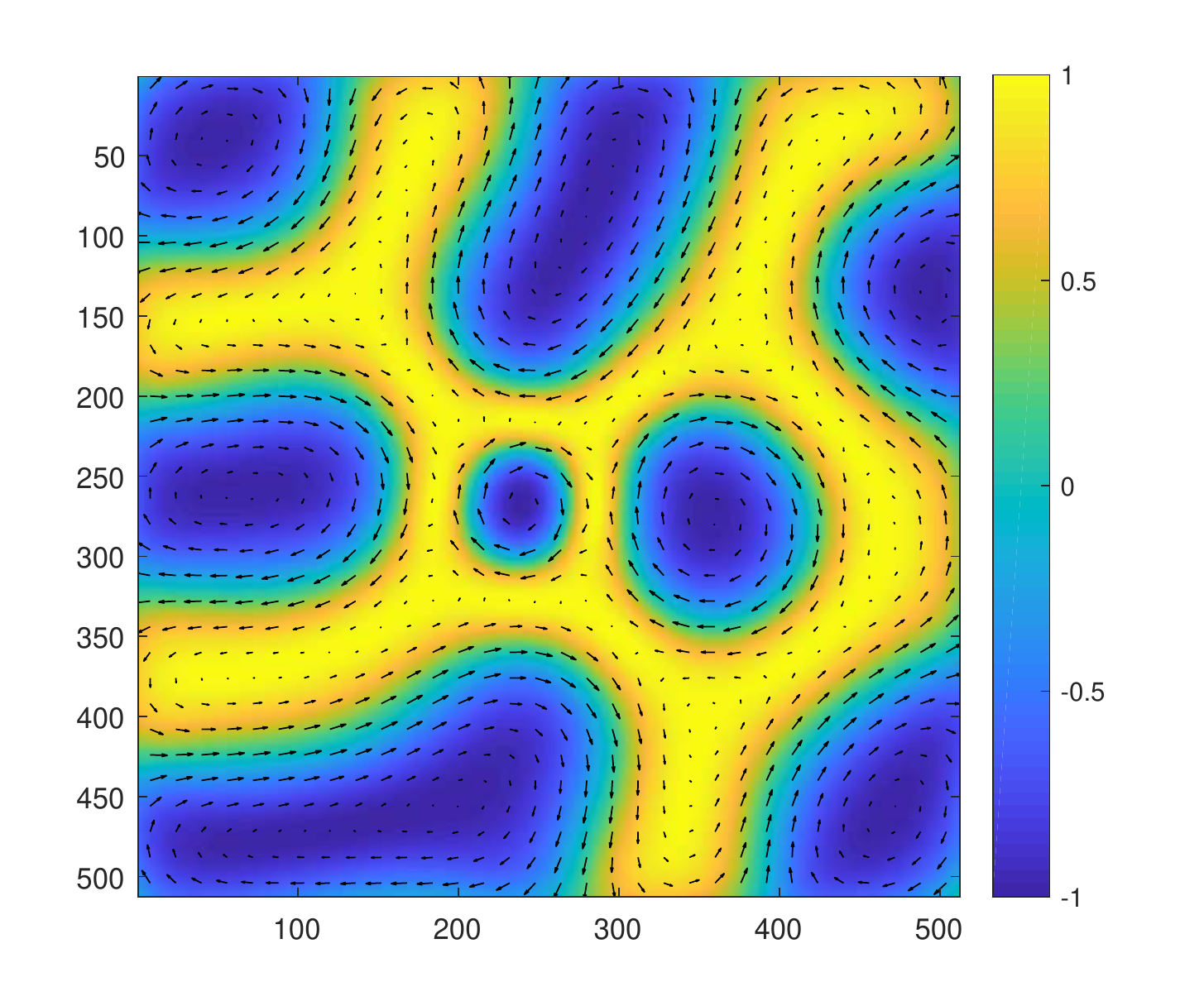}}\hfill
\subfloat[$D=0.02$,$\tau=120000$,$g=-1$,$m=1$,$\gamma\approx 4.7060$,$R_s\approx 84.6$]{\includegraphics[width=0.33\textwidth]{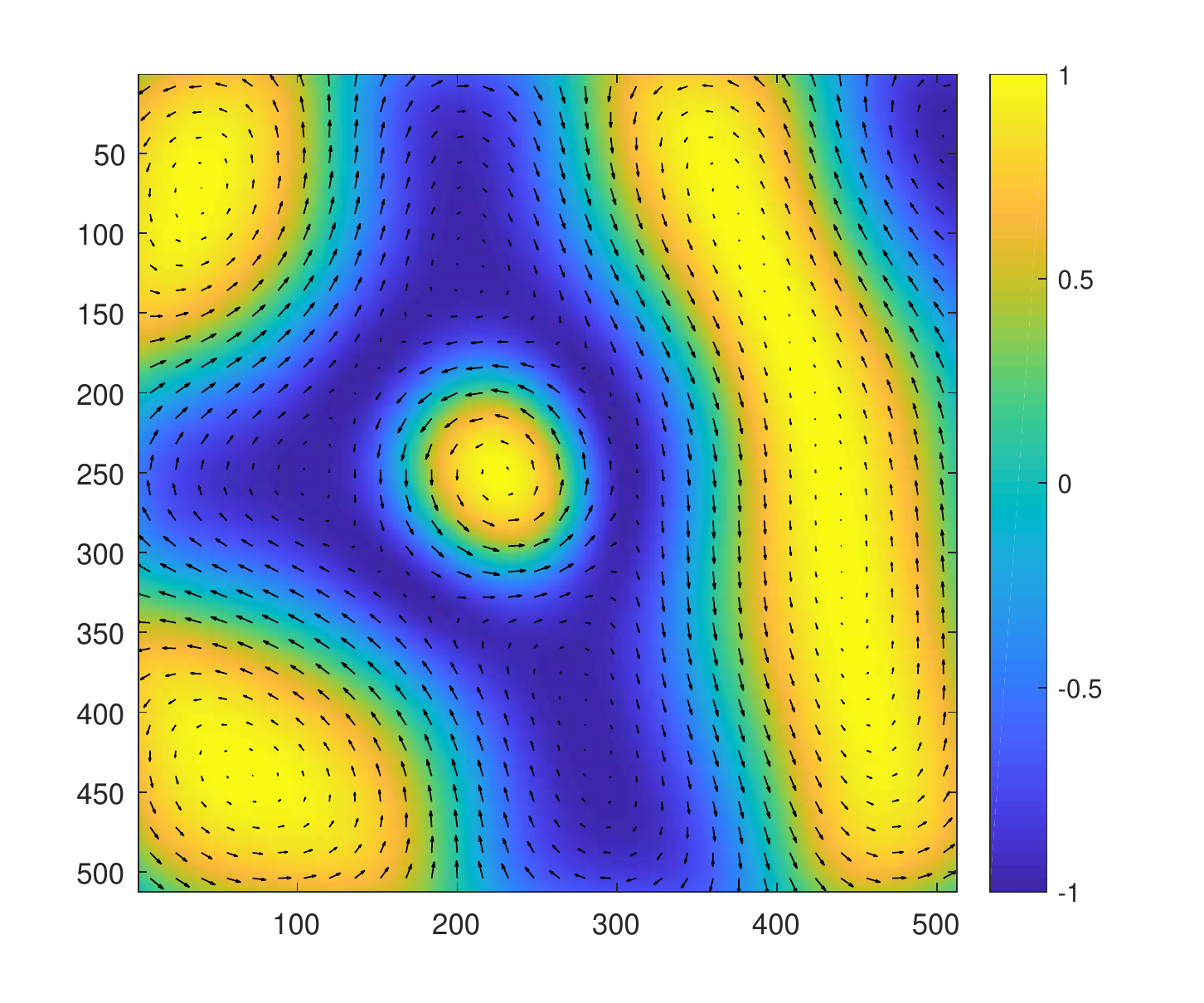}}\hfill
\subfloat[$D=0.01$,$\tau=140000$,$g=-1$,$m=1$,$\gamma\approx 4.7165$,$R_s\approx 182.0$]{\includegraphics[width=0.33\textwidth]{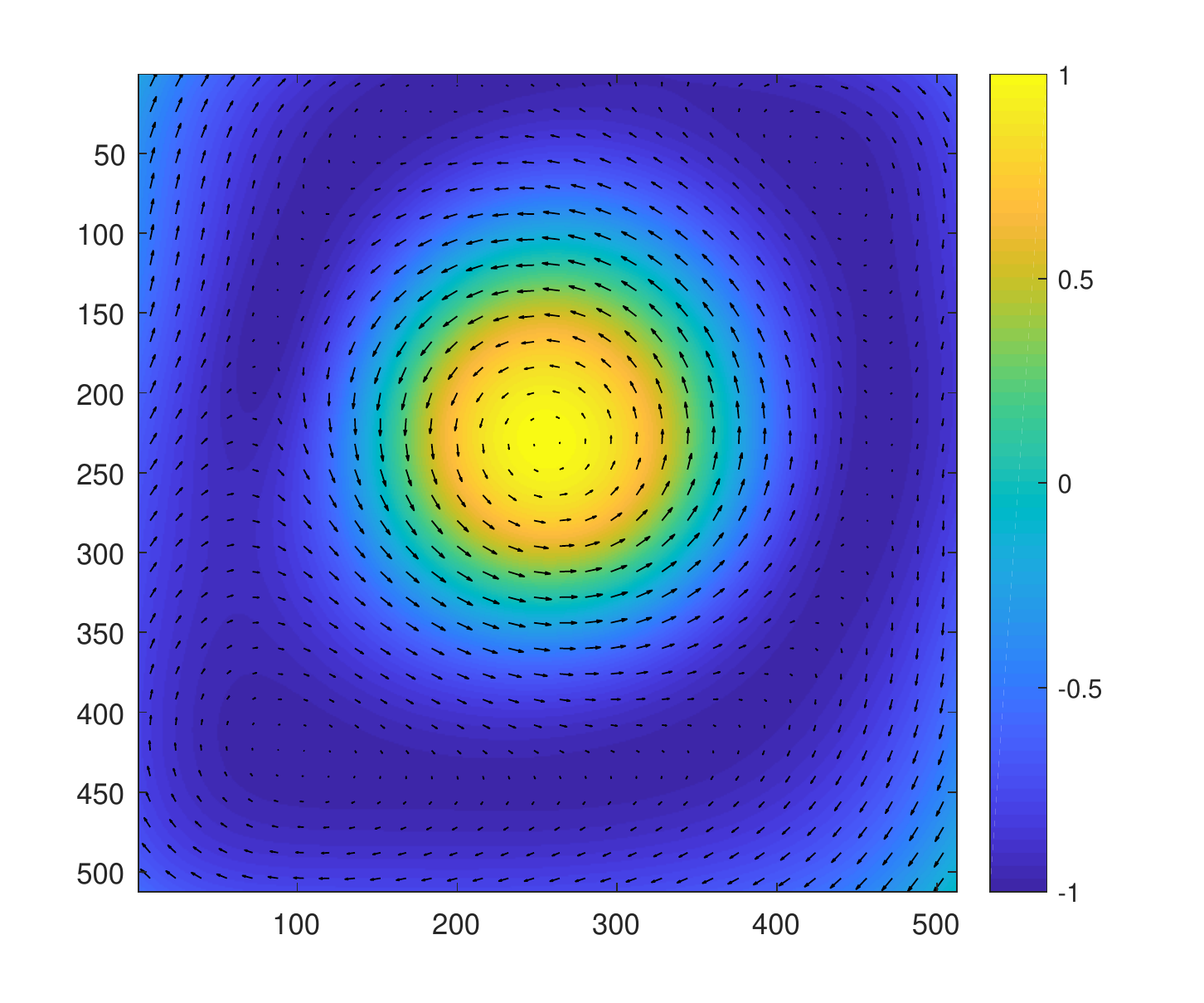}}\vfill
\caption{ The stable spin configuration ${\bf n}_r$ in the lattice  simulation  for $D=0.03, 0.02, 0.01$. The heat map represents the magnitude of  $n_z$, the arrow represents $(n_x,n_y)$. In (a), only   the skyrmion at the center is considered. }
\label{fig:results3}
\end{figure}

We have also studied  the case of $J_1<0$. Using $J_1=-0.5$, $J_2=0.0001$, $\alpha=0.2$, and $\Delta t=0.01$,  skyrmions  are generated in the simulation (Fig.~\ref{fig:results3}). In this case, the  range of the value of $D$ in which skyrmions can be generated is narrower than the case of $J_1=3$ and $J_2=0.001$. Similar to the case of $J_1>0$, the radius of the skyrmion  also increases with the decrease of $D$, and $\tilde{D}>0$ for all the skyrmions generated.

\subsection{\label{sec:4.2} Skyrmion generation  from helical  phase }

We have also studied a more realistic situation. Starting with the  helical phase, we suddenly switch on the pinning, by substituting the constant $J$ with an inhomogeneous $J(\rho)$. In   previous works, $D/J\approx 0.09\sim 0.5$~\cite{Tchoe,pin,unitJ,alpha3,bilayer,LLG2,discreteH,dvalue}, therefore  we use $D/J_0=0.08,0.1,0.2,0.3,0.5$.   In absence of an external magnetic field and with constant $J$,  the initial state is prepared by relaxing from a saturated state with ${\bf n}_{\bf r}={\bf e}_z$ until the stability is reached. Thus the helical phase is obtained. Then we substitute the constant $J$ with inhomogeneous $J(\rho)$ at $\tau = 0$ and continue the simulation. For $D=0.08$, we use $\alpha = 0.2$ and $\Delta t = 0.01$. For $D=0.1,0.2,0.3,0.5$, we use $\alpha =0.1$ and $\Delta t = 0.002$.

\begin{figure}
\subfloat[$\tau=0$]{\includegraphics[width=0.33\textwidth]{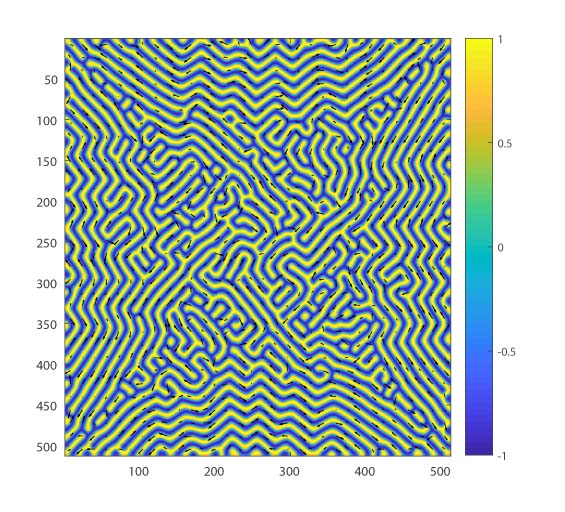}}\hfill
\subfloat[$\tau=40$]{\includegraphics[width=0.33\textwidth]{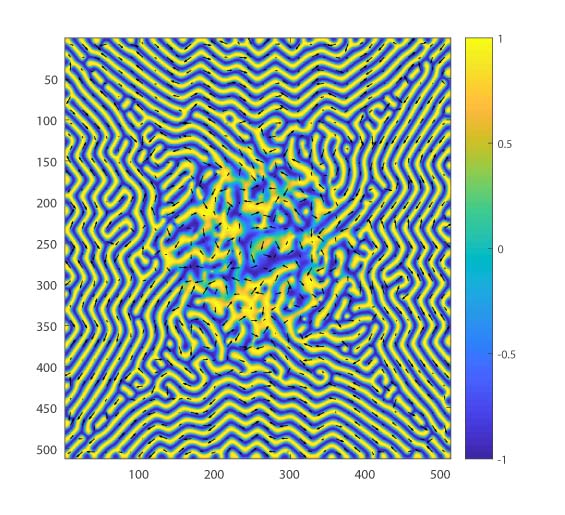}}\hfill
\subfloat[$\tau=80$]{\includegraphics[width=0.33\textwidth]{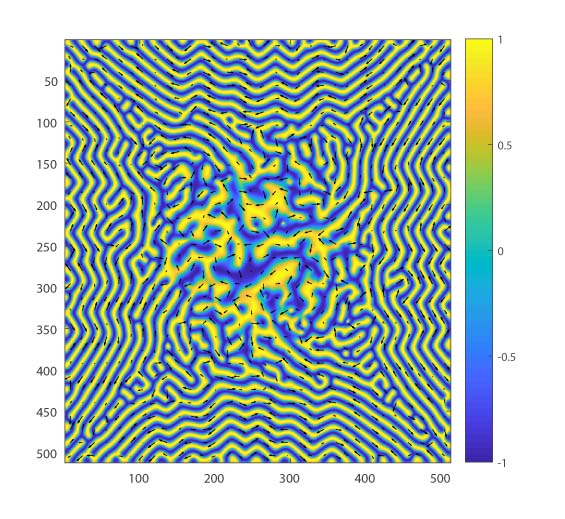}}\vfill
\subfloat[$\tau=160$]{\includegraphics[width=0.33\textwidth]{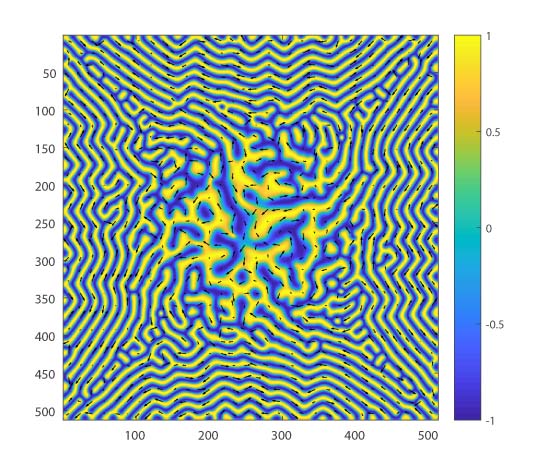}}\hfill
\subfloat[$\tau=400$]{\includegraphics[width=0.33\textwidth]{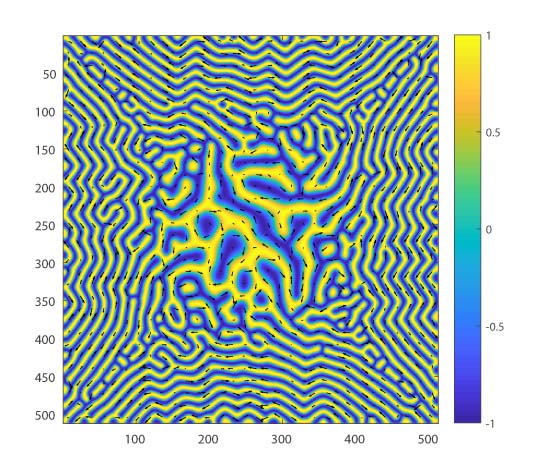}}\hfill
\subfloat[$\tau=10000$]{\includegraphics[width=0.33\textwidth]{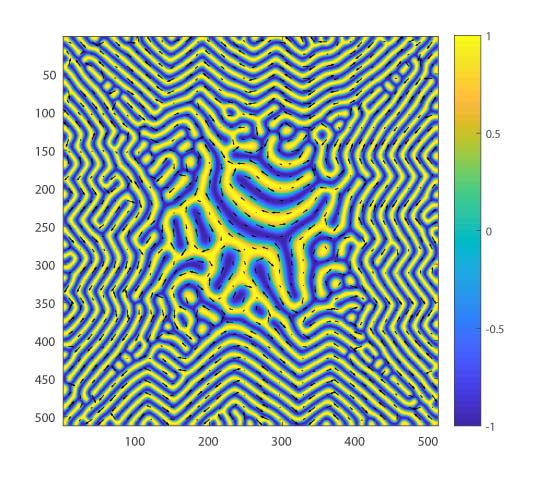}}\vfill
\caption{Simulation starting from helical phases with $D=0.5$. Pinning is switched on by suddenly substituting the constant $J$ with $J(\rho)$, with $J_1=3,J_2=0.0001$. With inhomogeneous $J(\rho)$, the stripes  grow  while turbulence is created and small skyrmions are generated, some of which  become stable  finally. }
\label{fig:impD05}
\end{figure}

The results for $D=0.5$, $J_1=3$ and $J_2=0.0001$ are shown in Fig.~\ref{fig:impD05}. The strip width increases with the decrease of  $D/J$. So the  stripes are wider near the pinning center, where  $D/J$ is smaller.  The stripes start to grow at $\tau = 0$. Then a kind of turbulence is created, and small skyrmions are generated, some of which finally become stable.

\begin{figure}
\subfloat[$D=0.08,\tau=0$]{\includegraphics[width=0.25\textwidth]{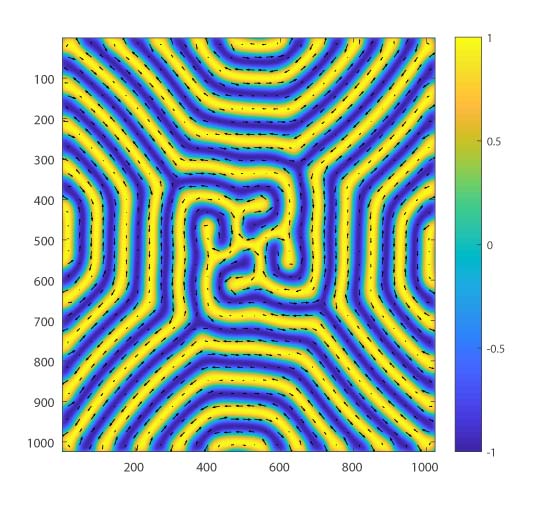}}\hfill
\subfloat[$D=0.08,\tau=50000$]{\includegraphics[width=0.25\textwidth]{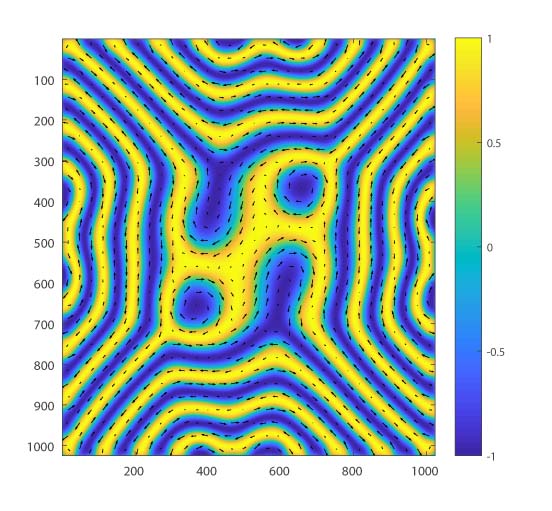}}\hfill
\subfloat[$D=0.1,\tau=0$]{\includegraphics[width=0.25\textwidth]{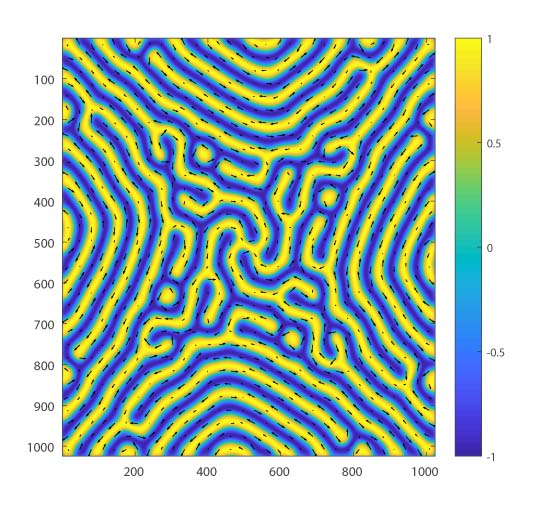}}\hfill
\subfloat[$D=0.1,\tau=12000$]{\includegraphics[width=0.25\textwidth]{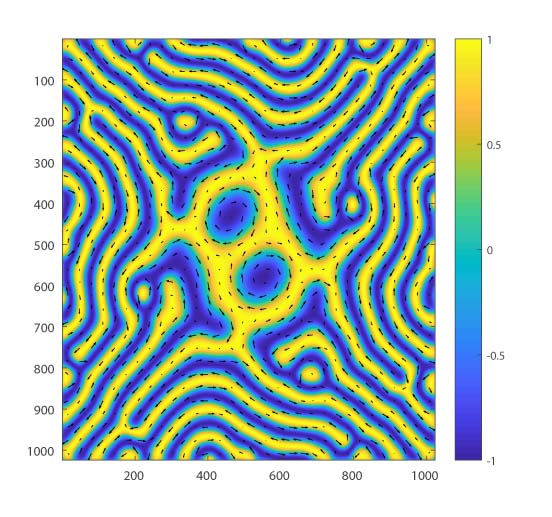}}\vfill
\subfloat[$D=0.2,\tau=0$]{\includegraphics[width=0.25\textwidth]{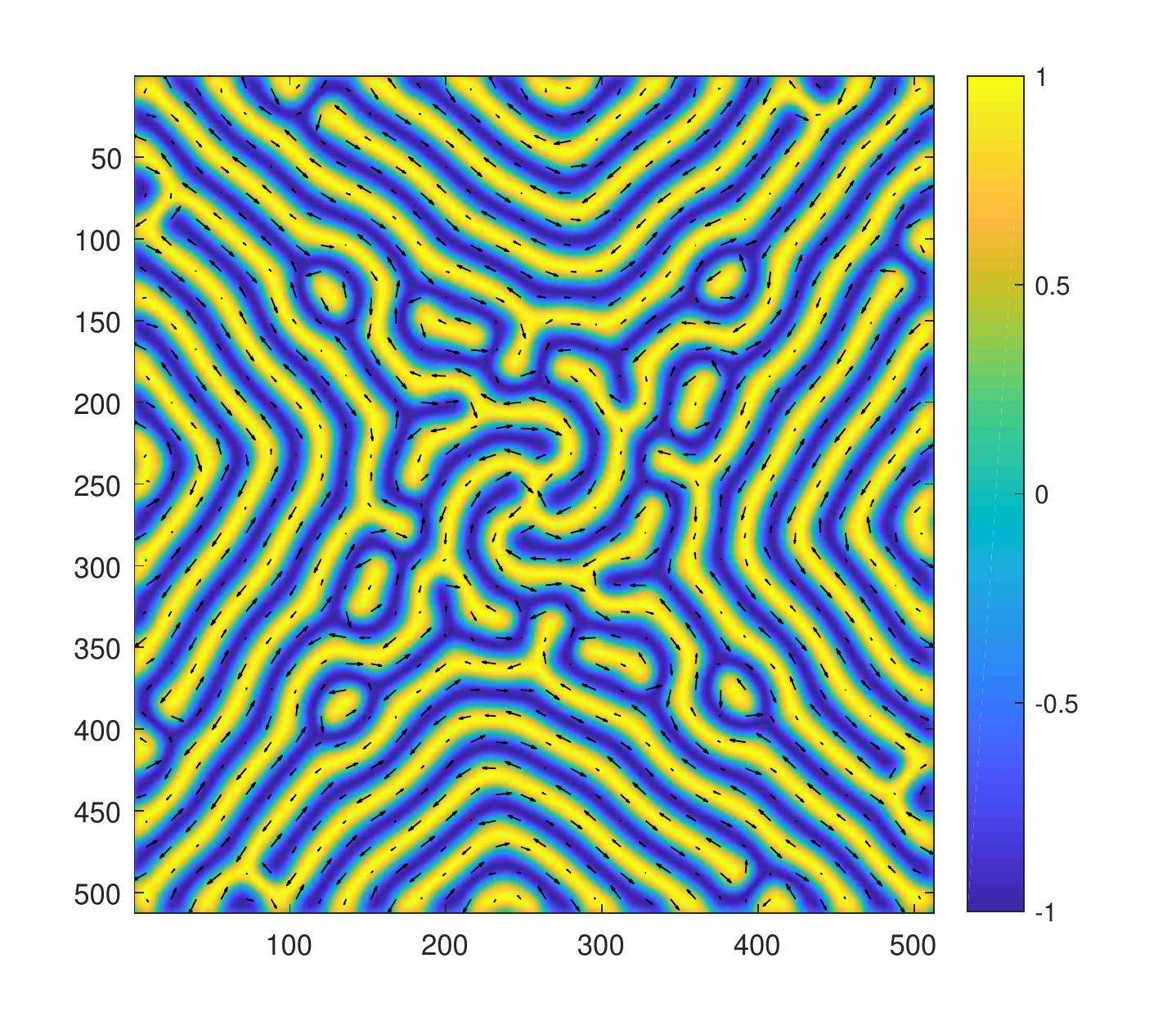}}\hfill
\subfloat[$D=0.2,\tau=10000$]{\includegraphics[width=0.25\textwidth]{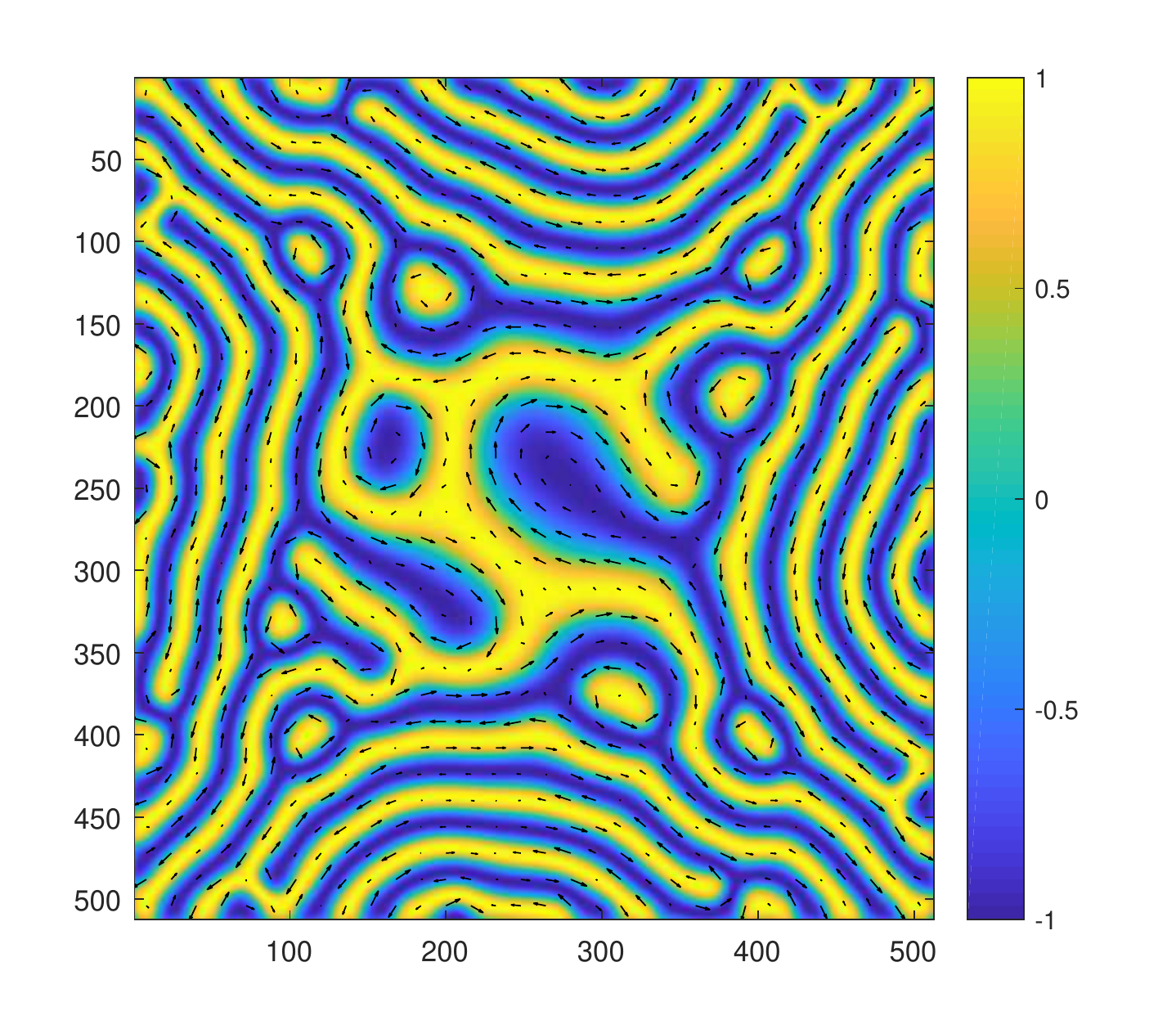}}\hfill
\subfloat[$D=0.3,\tau=0$]{\includegraphics[width=0.25\textwidth]{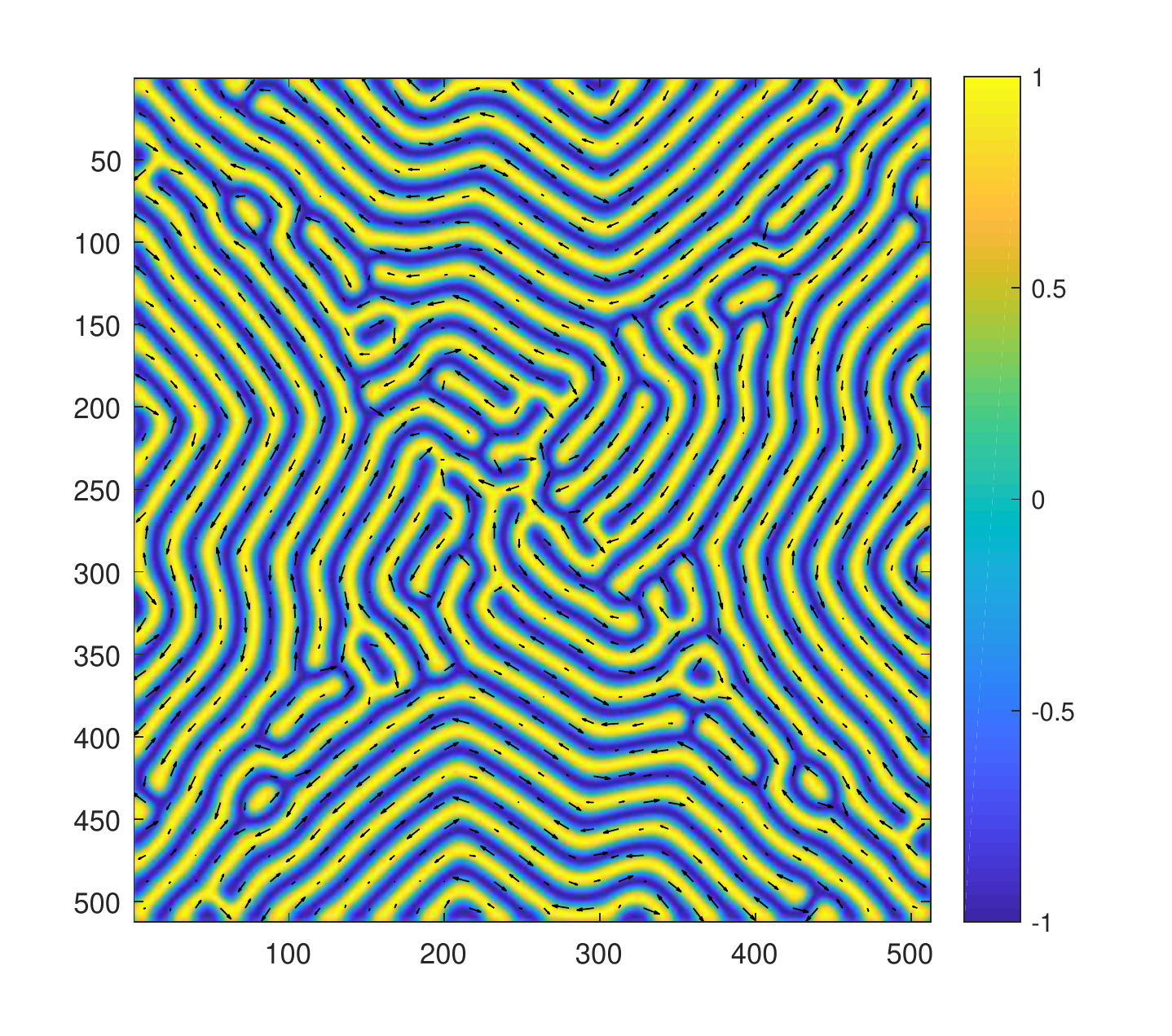}}\hfill
\subfloat[$D=0.3,\tau=10000$]{\includegraphics[width=0.25\textwidth]{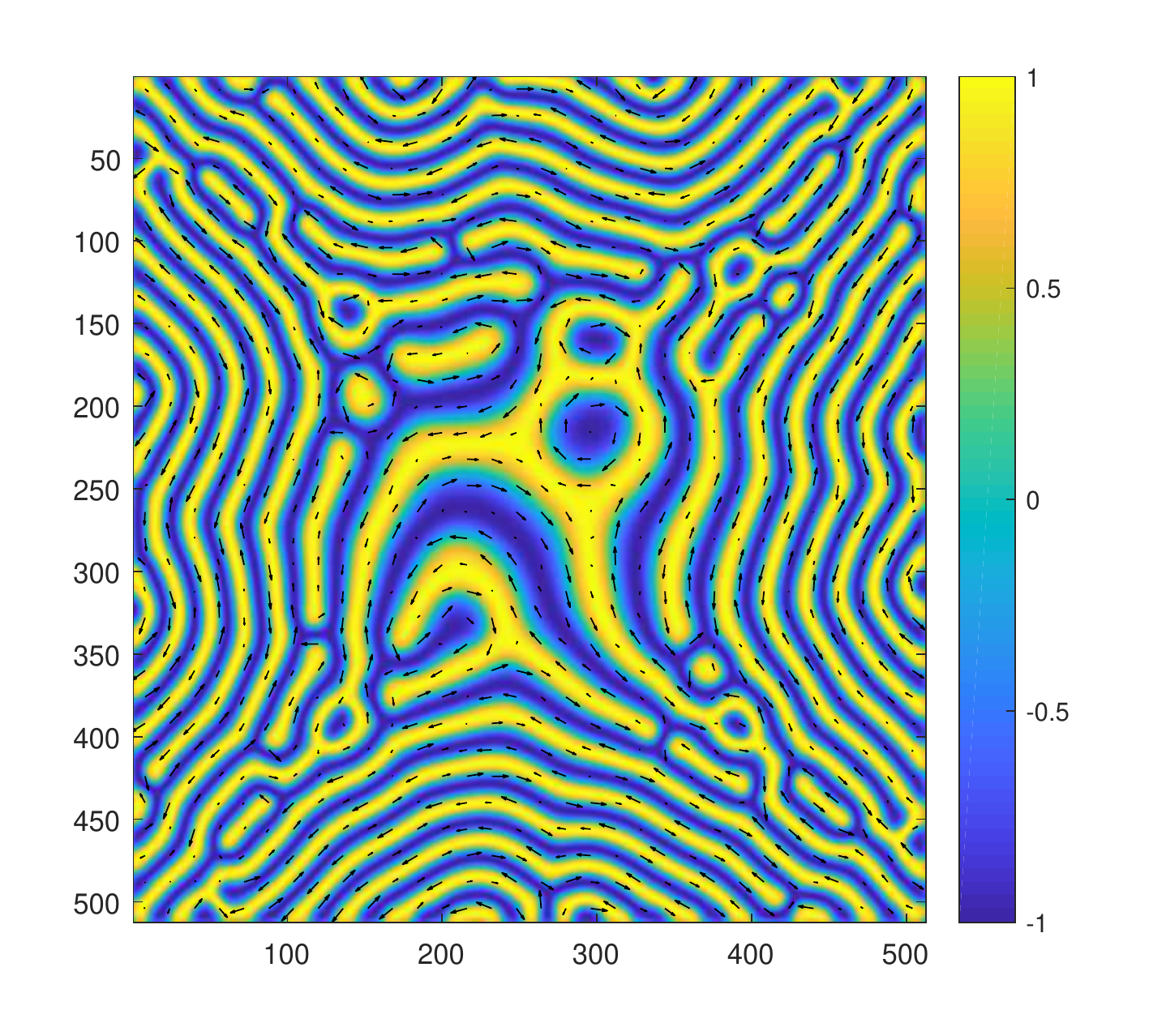}}\vfill
\caption{Simulation starting from helical phases. Pinning is applied by suddenly substituting the constant $J$ with $J(\rho)$. For $D=0.2,0.3$, $J_1=3,J_2=0.0001$. For $D=0.08,0.1$, we run the simulation on a $1024\times 1024$ lattice, with $J_1=3,J_2=0.000025$ and the pinning center locating at $(512,512)$. In each figure, skyrmions are  generated in the pinning region, though  not at the pinning center. Some skyrmions originate  from short stripes close to skyrmions, but new skyrmions are also generated.}
\label{fig:imps}
\end{figure}

The cases for $D=0.08,0.1,0.2,0.3$ are shown in Fig.~\ref{fig:imps}. For $D=0.2, 0.3$, we still use $J_2=0.0001$, as in the case of  $D=0.5$.  The size of the pinning should be much larger than the width of the stripes.  Hence for smaller values of $D$, $J_2$ is chosen to be smaller, consequently the lattice is set to be larger. Therefore, for $D=0.08, 0.1$, we run the simulation on a $1024\times 1024$ lattice. We choose $J_2=0.000025$ and set the pinning  center to be at $(512,512)$. For all these cases,   skyrmions  are  generated in the  pinning region (Fig.~\ref{fig:imps}), although they are not at  the pinning  center,  as in the cases  starting with randomized initial state.

\begin{figure}
\subfloat[$\tau=0$]{\includegraphics[width=0.5\textwidth]{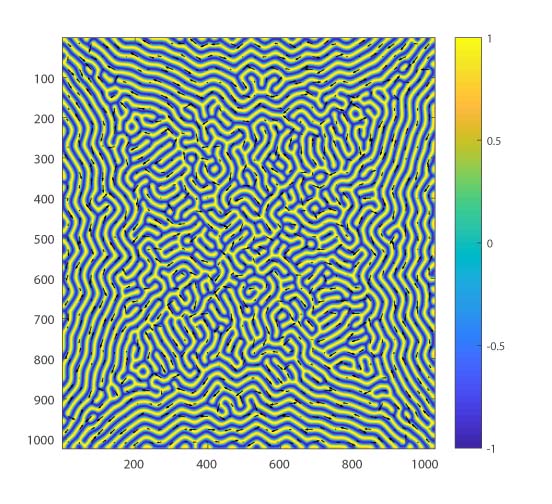}}\hfill
\subfloat[$\tau=3000$]{\includegraphics[width=0.5\textwidth]{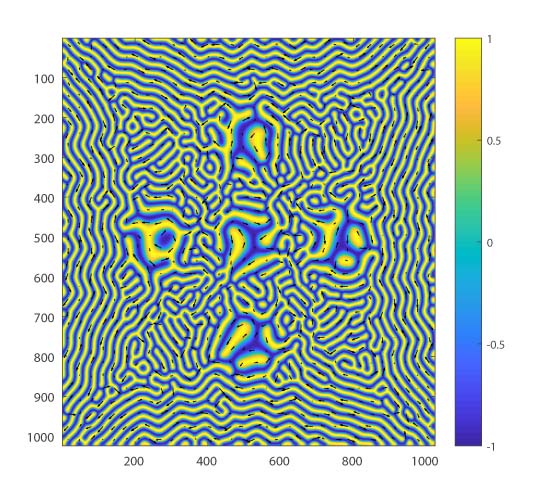}}\vfill
\caption{Simulation starting from a helical phase,  with $D=0.3$. There are $5$ pinning centers at the sites $(512,512)$, $(256,512)$, $(768,512)$, $(512,256)$ and $(512,768)$, all with  $J_2=0.00025$. The pinning radii are smaller than the those in Fig.~\ref{fig:imps}.~(c) and (d). The skyrmions are generated in the pinning  regions near $(256,512)$, $(768,512)$, $(512,256)$ and $(512,768)$, but not near $(512,512)$, where the spin structure is not topologically separated from the other stripes. }
\label{fig:impD03}
\end{figure}

For $D=0.3$, we run the simulation  on a $1024\times 1024$ lattice,  with $5$ pinning centers locating  at  sites ${\bf r}_1= (512,512)$, ${\bf r}_2=(256,512)$, ${\bf r}_3=(768,512)$, ${\bf r}_4=(512,256)$ and ${\bf r}_5=(512,768)$, respectively,
\begin{equation}
\begin{split}
&J({\bf r})=1+3\sum _{i=1}^5 \exp \left(-0.00025 ({\bf r} -{\bf r} _i)^2\right),
\end{split}
\label{eq.4.6}
\end{equation}
with $J_2=0.00025$.  The result (Fig.~\ref{fig:impD03}) shows that skyrmions are generated near some but not all  of the  pinning centers.

We also find that, for  $J_1<0$, skyrmions cannot be generated by using this method.

\subsection{\label{sec:4.3}  Skyrmion generation with   both \texorpdfstring{$J$}{J} and \texorpdfstring{$D$}{D}   inhomogeneous }

Although the origins of the ferromagnetic exchange and DM interaction are different, it is difficult to vary $J$ while keeping $D$ constant in real experiments. We also simulate a scenario in which  both $J$ and $D$  inhomogeneous while $D/J$ constant, that is, $D$ has the same $\rho$-dependence as $J$~\cite{pin}.  When $D/J$ is homogeneous, the widths of the stripes are not changed.  In this case, skyrmions cannot be generated from helical phases. However,  skyrmions can still be generated at the pinning center through the  pinning effect with the aid of the boundary condition, in the following way.

\begin{figure}
\includegraphics[width=0.8\textwidth]{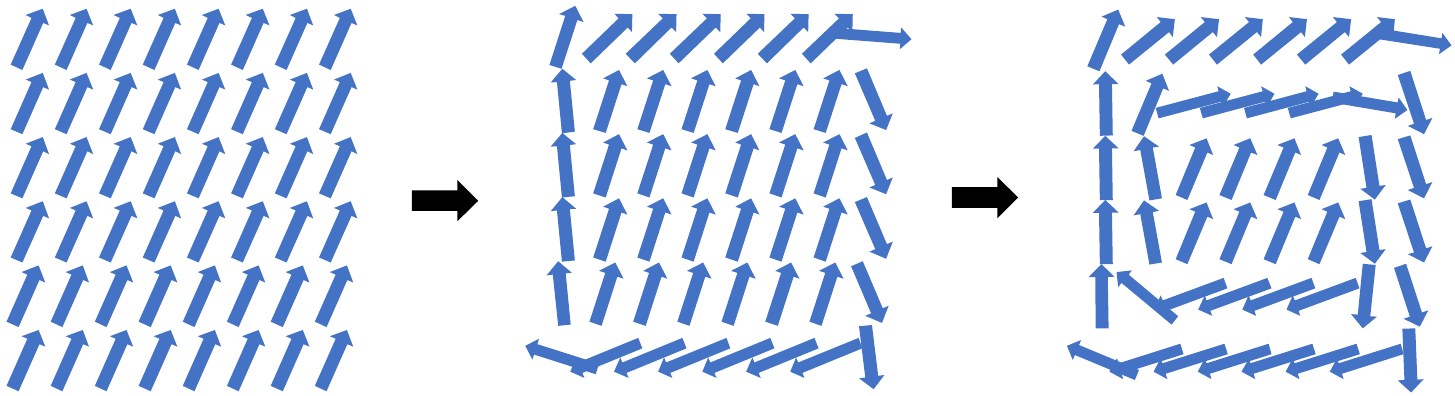}
\caption{When  the simulation starts from ${\bf n}_{\bf r}={\bf e}_z$, the boundary plays an important role. With DM interaction, ${\bf n}_{\bf r}$ at a site tends to tilt to its neighbours. If $D$ is homogenous, the tilt starts from the boundary.}
\label{fig:tilt}
\end{figure}

In our simulation,  when an external magnetic field is applied, ${\bf n}_{\bf r}$ can be saturated to ${\bf n}_{\rm r}={\bf e}_z$ no matter whether pinning is presented. If we suddenly switch off the external magnetic field and start the simulation from a saturated initial configuration, the boundary condition  plays an important role. Because of DM interaction, ${\bf n}_{\bf r}$   tends to tilt to  its neighbours. If $D$ is homogenous, the tilt starts from the sites on the boundary (Fig.~\ref{fig:tilt}). When $D$ is inhomogeneous, the tilt starts from both the boundary and the sites with inhomogeneous $D$. This can be understood by  using LLG in Eq.~(\ref{eq.4.2}), which can be rewritten as
\begin{equation}
\begin{split}
&\frac{d{\bf n}}{dt}=\frac{1}{1+\alpha^2}\left({\bf N}+\alpha {\bf N}\times {\bf n}\right),
\end{split}
\label{eq.4.7}
\end{equation}
with ${\bf N}={\bf B}_{\rm eff}\times {\bf n}$. When ${\bf n}_{\bf r}={\bf e}_z$ and $D$ is homogeneous, ${\bf B}_{\rm eff}\propto {\bf n}_{\bf r}$ and ${\bf N}=0$. However, when $D$ is inhomogeneous, ${\bf N}\neq 0$, therefore ${\bf n}_{\bf r}$ of the sites at the boundary or at the sites with inhomogeneous $D$  start to tilt first.

\begin{figure}
\subfloat[$\tau=20$]{\includegraphics[width=0.33\textwidth]{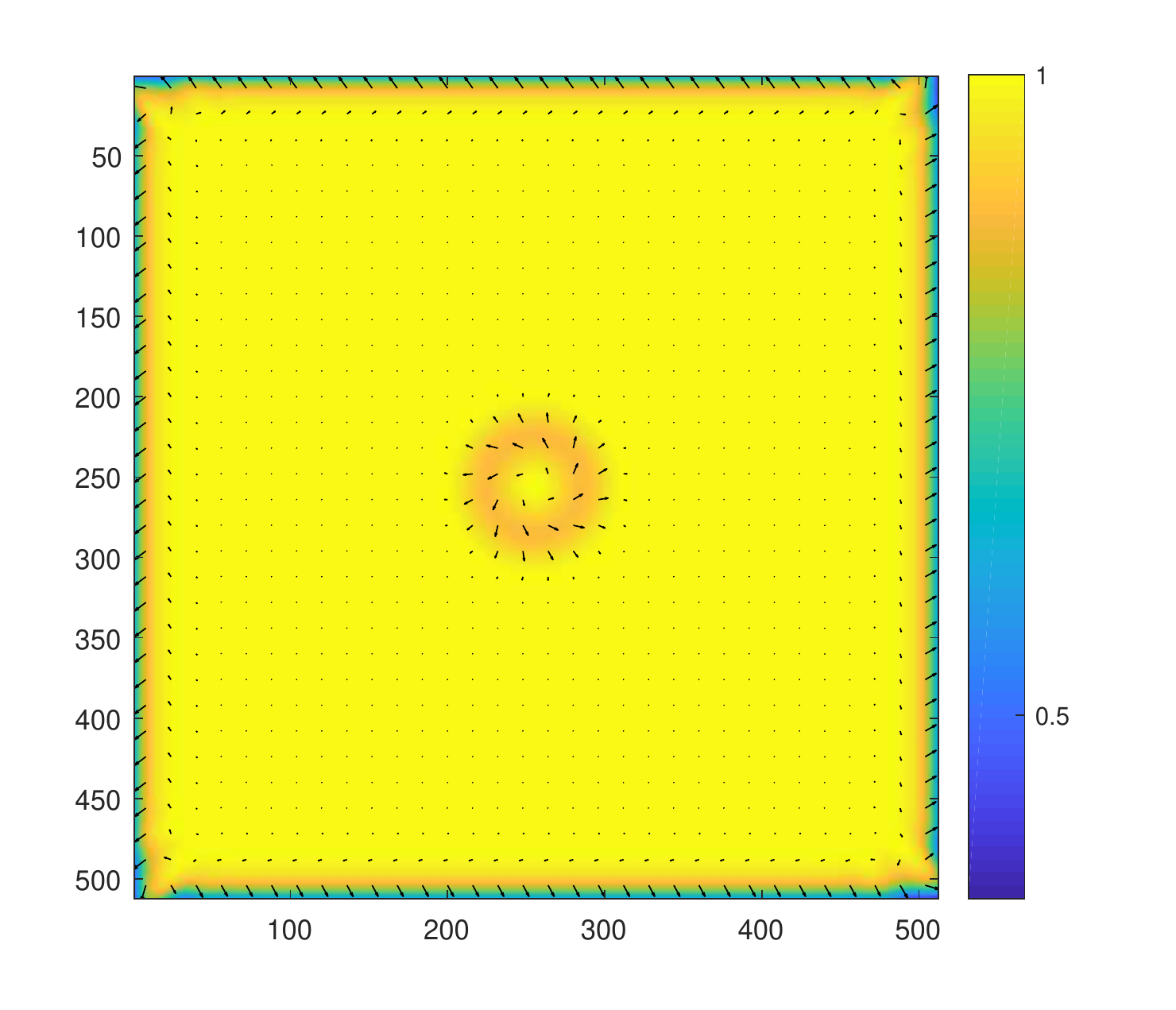}}\hfill
\subfloat[$\tau=200$]{\includegraphics[width=0.33\textwidth]{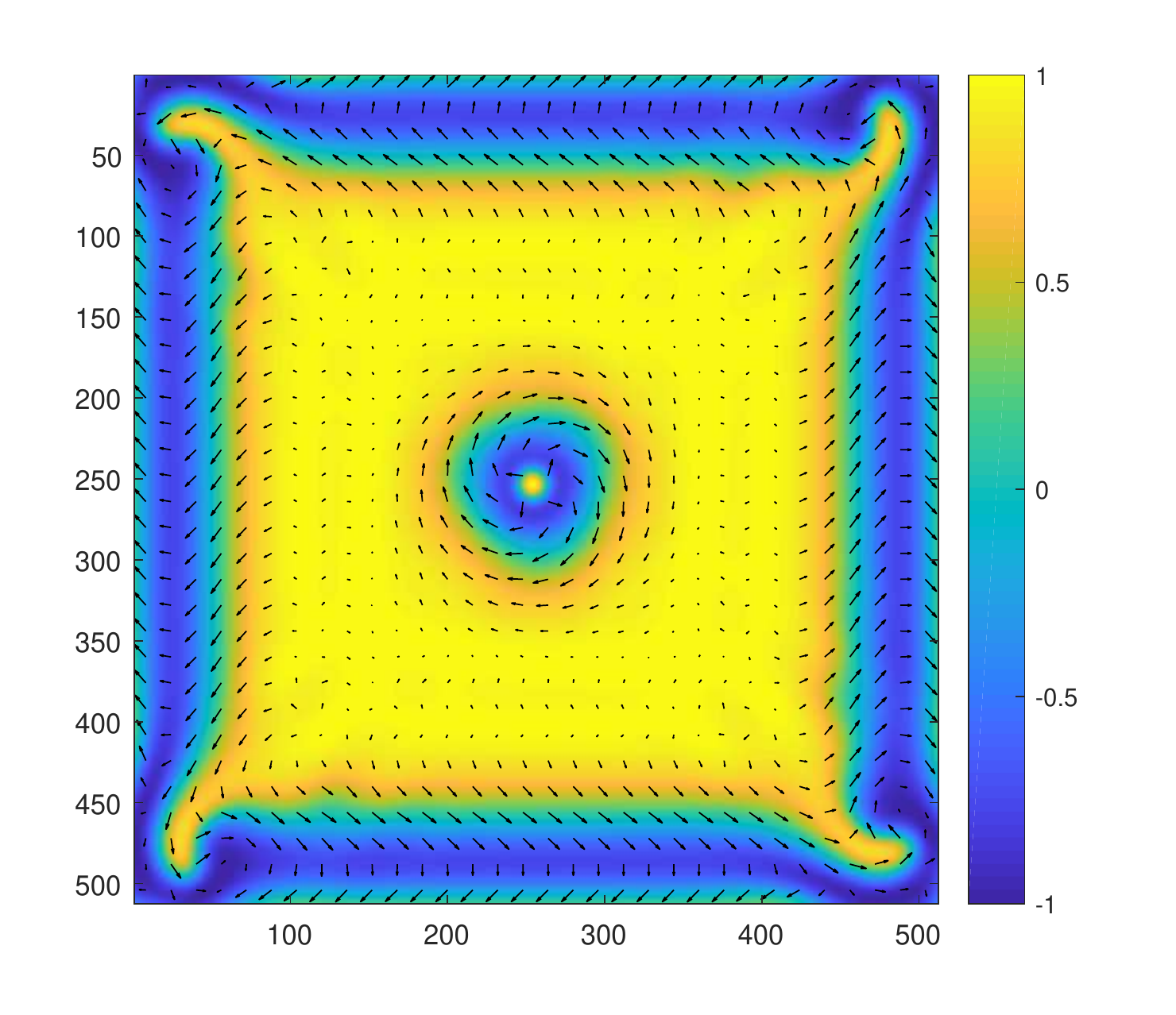}}\hfill
\subfloat[$\tau=400$]{\includegraphics[width=0.33\textwidth]{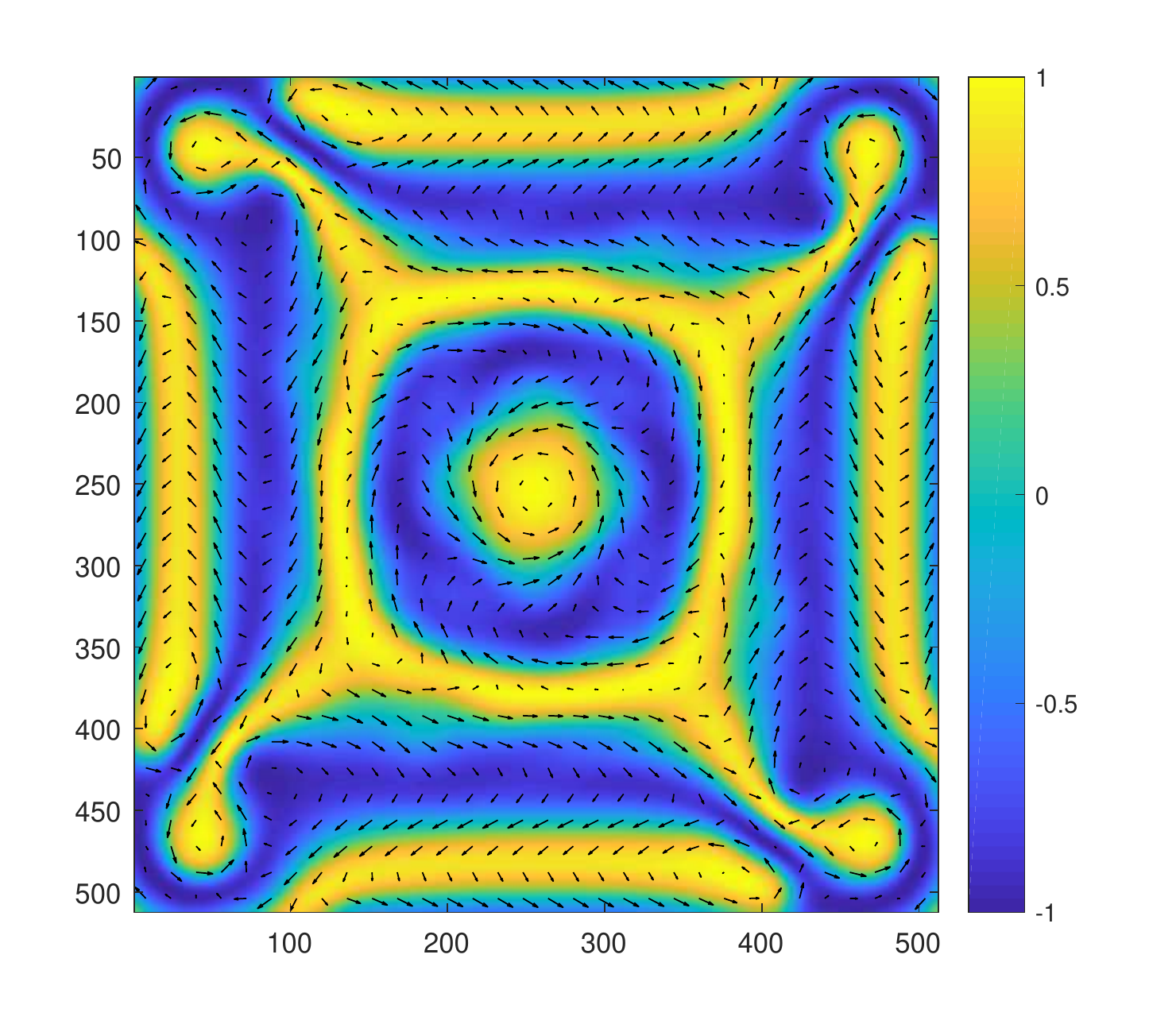}}\vfill
\subfloat[$\tau=600$]{\includegraphics[width=0.33\textwidth]{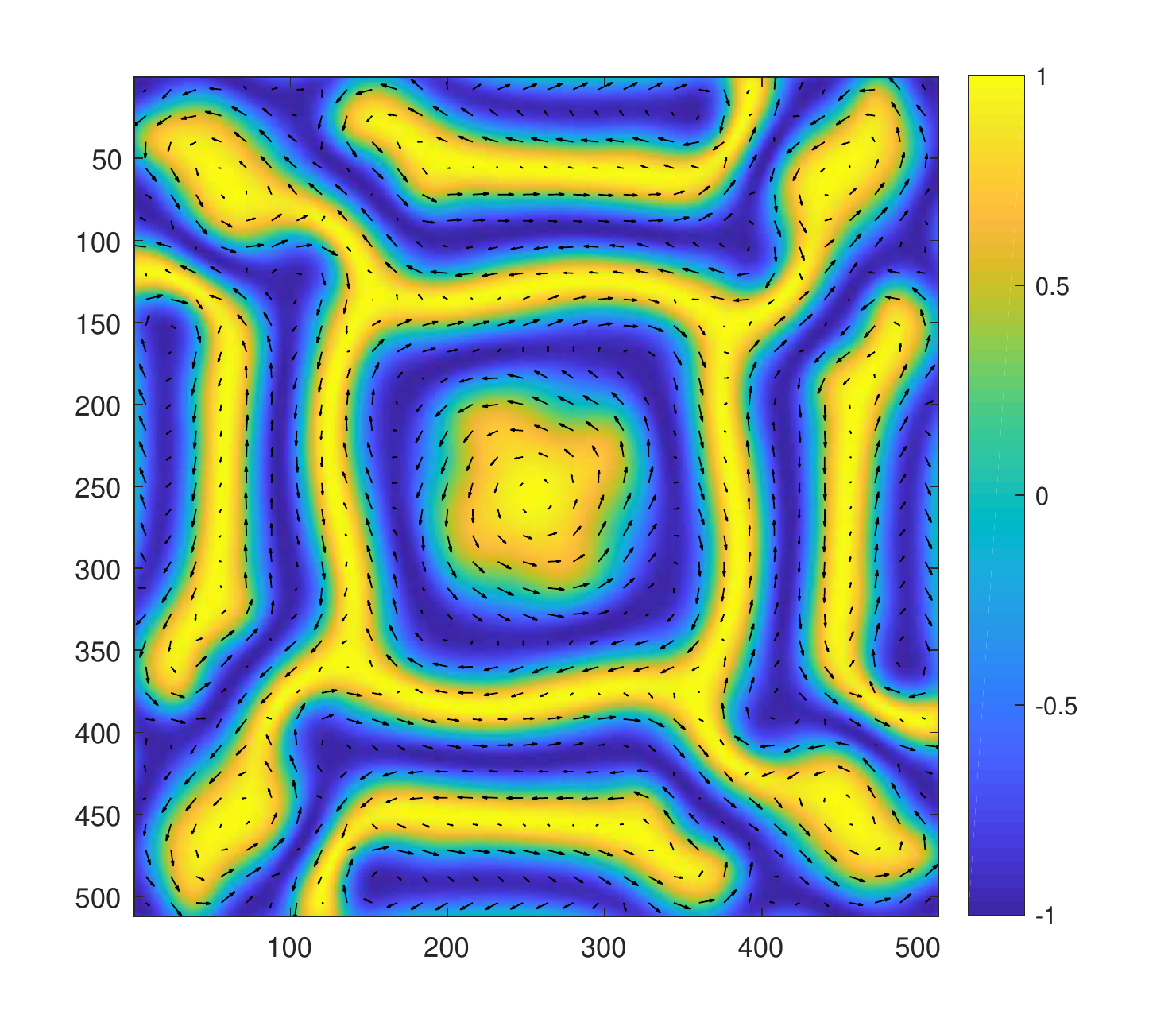}}\hfill
\subfloat[$\tau=800$]{\includegraphics[width=0.33\textwidth]{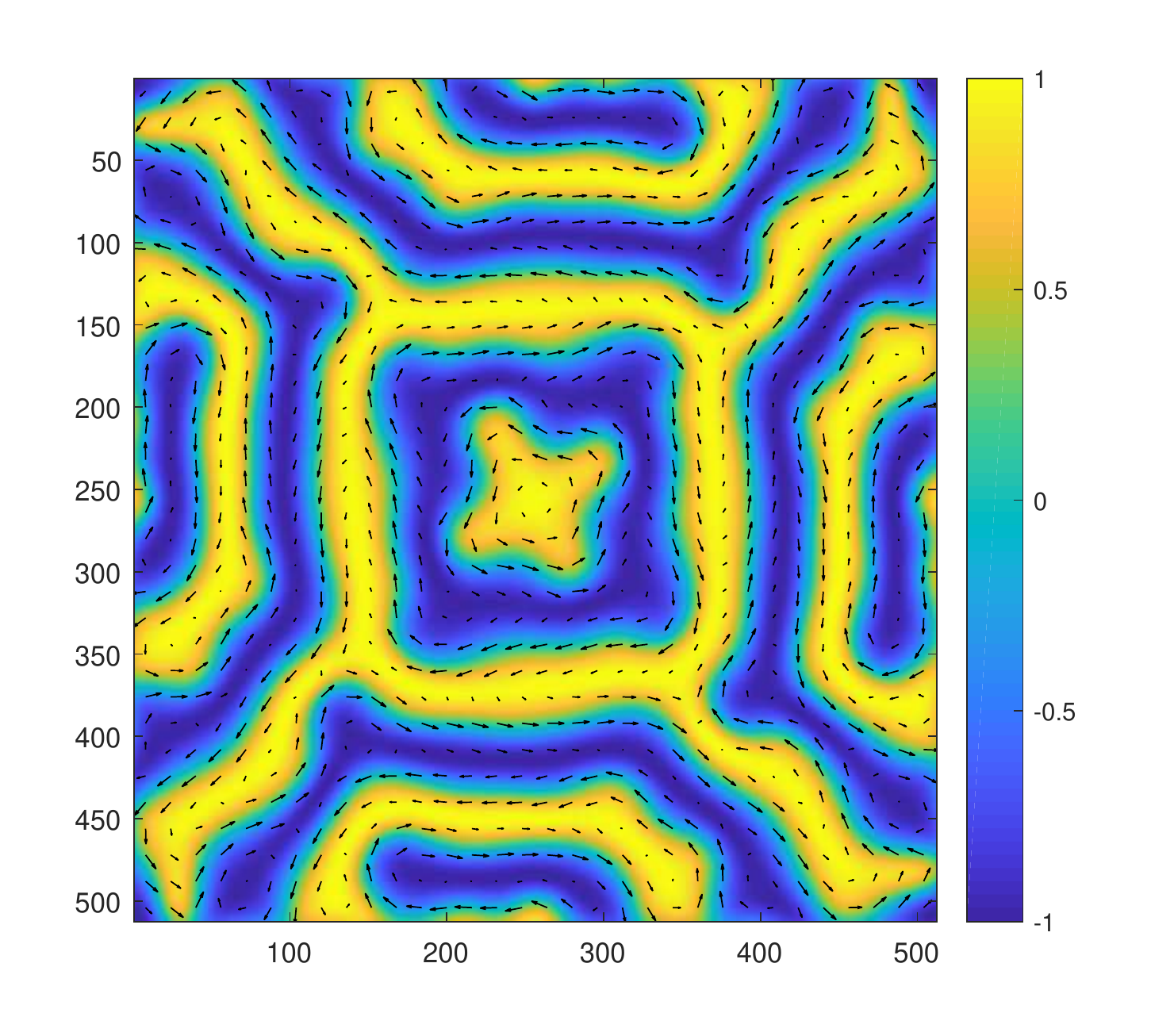}}\hfill
\subfloat[$\tau=5000$]{\includegraphics[width=0.33\textwidth]{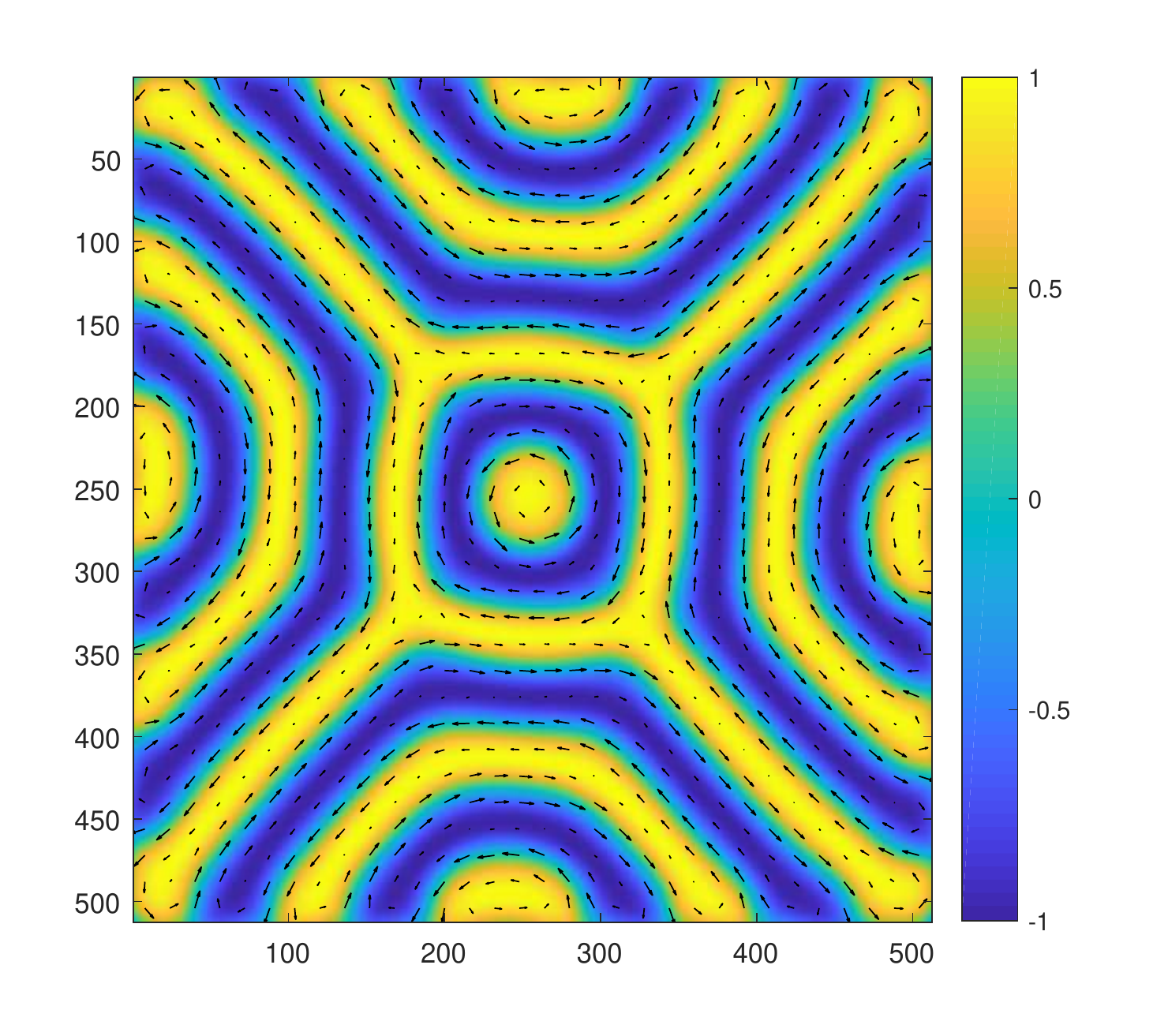}}\vfill
\caption{Simulation for $D/J=0.08$, $J_1=3$, $J_2=0.001$. Because of the inhomogeneous $D$, a ring domain wall is generated around the pinning at about $\tau =20$ and keeps shrinking to the center of the ring. The center of the ring is a skyrmion. The ring keeps shrinking until about $\tau =200$, then the skyrmion at the center starts  to grow until it is constrained by the domain walls generated from the boundary.}
\label{fig:satD008}
\end{figure}

For $D/J=0.08$, we use $J_1=3$, $J_2=0.001$, $\alpha=0.2$, $\Delta t=0.002$. The result is shown in Fig.~\ref{fig:satD008}. Because of the inhomogeneous $D$, a ring domain wall is generated around the pinning and keeps shrinking to the center of the ring. The center of the ring is a skyrmion. The ring keeps shrinking until about $\tau = 200$, then the skyrmion at the center starts to grow until it is constrained by the domain walls generated from the boundary.

\begin{figure}
\subfloat[$\tau=200$]{\includegraphics[width=0.33\textwidth]{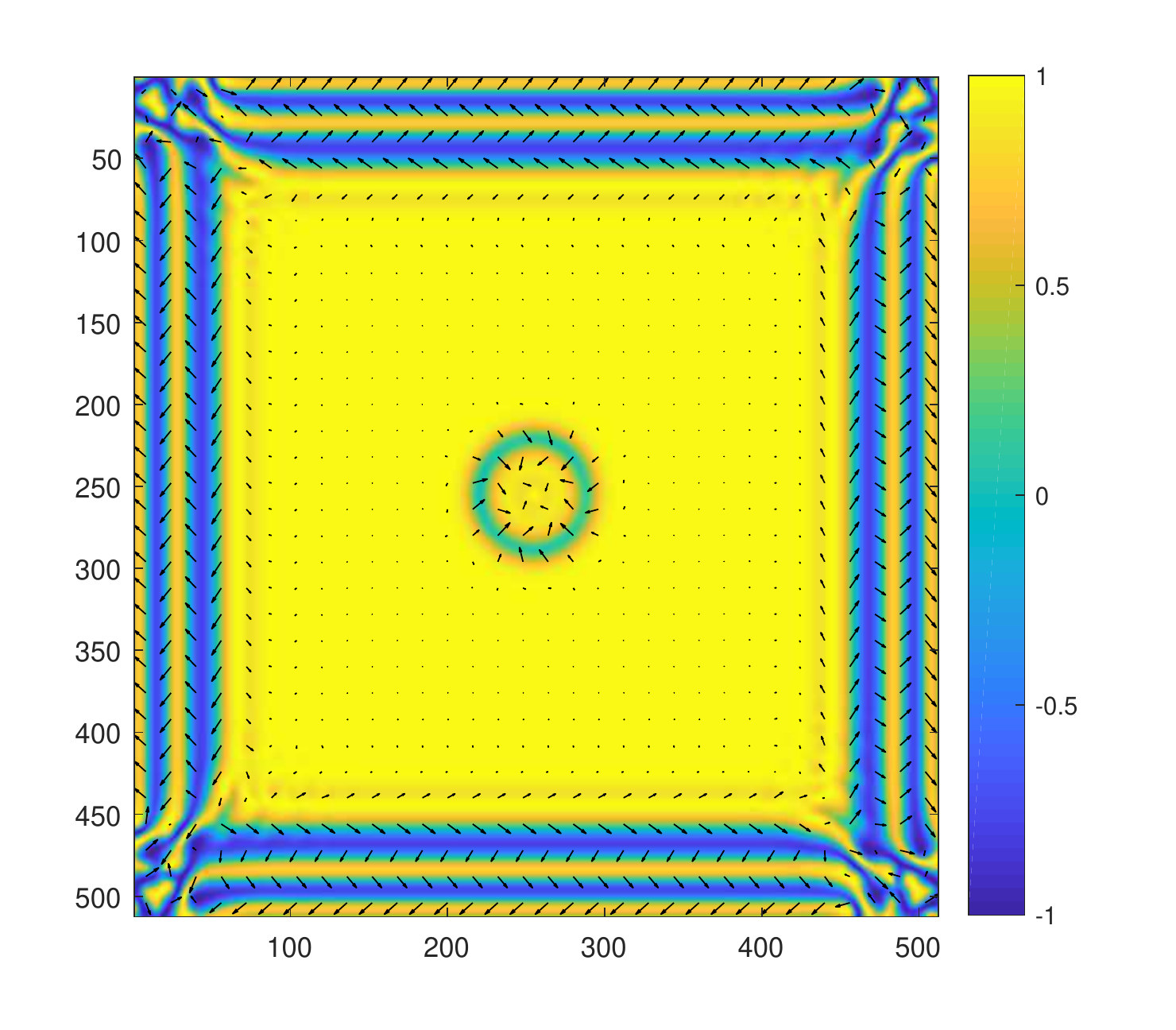}}\hfill
\subfloat[$\tau=400$]{\includegraphics[width=0.33\textwidth]{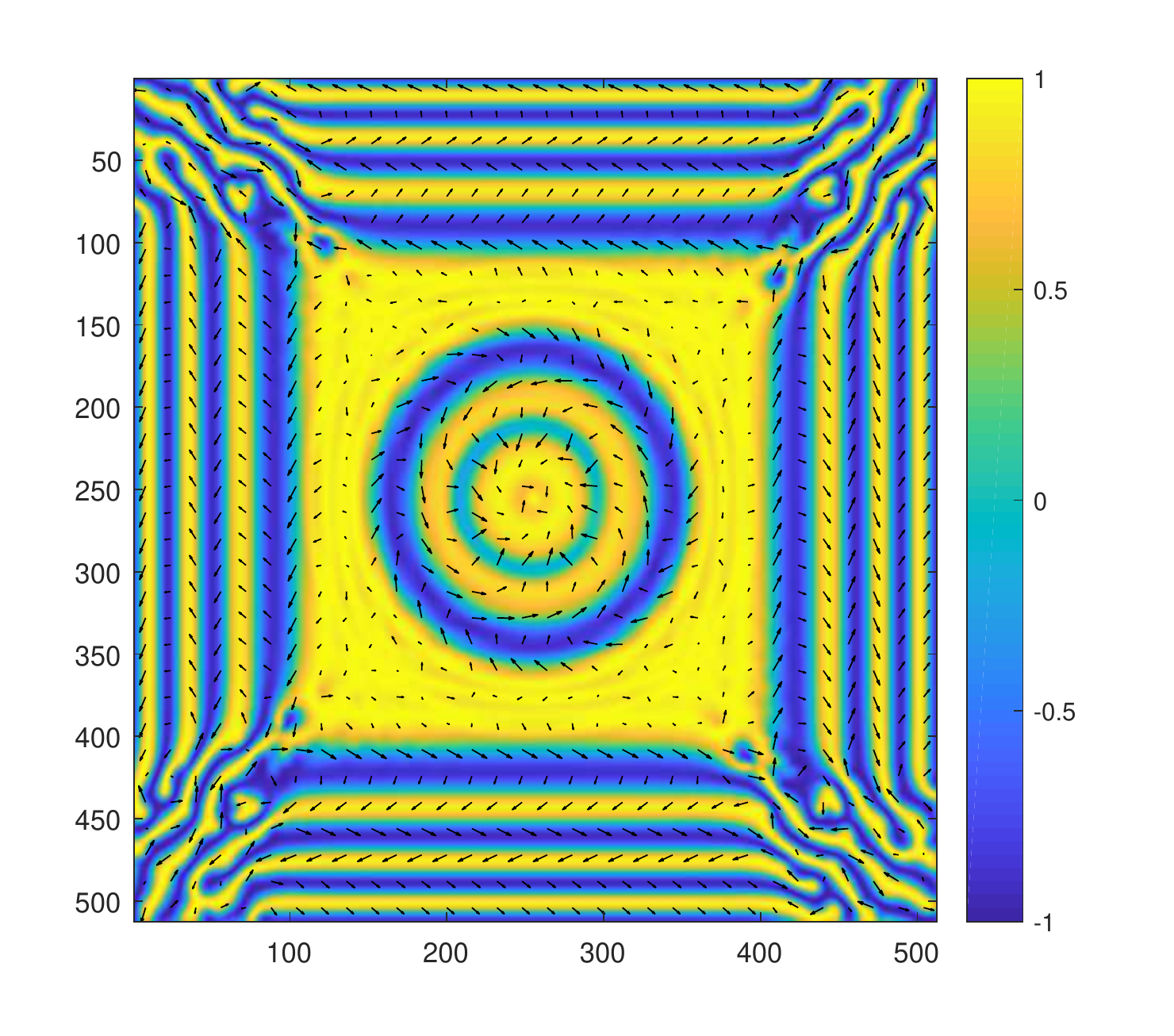}}\hfill
\subfloat[$\tau=600$]{\includegraphics[width=0.33\textwidth]{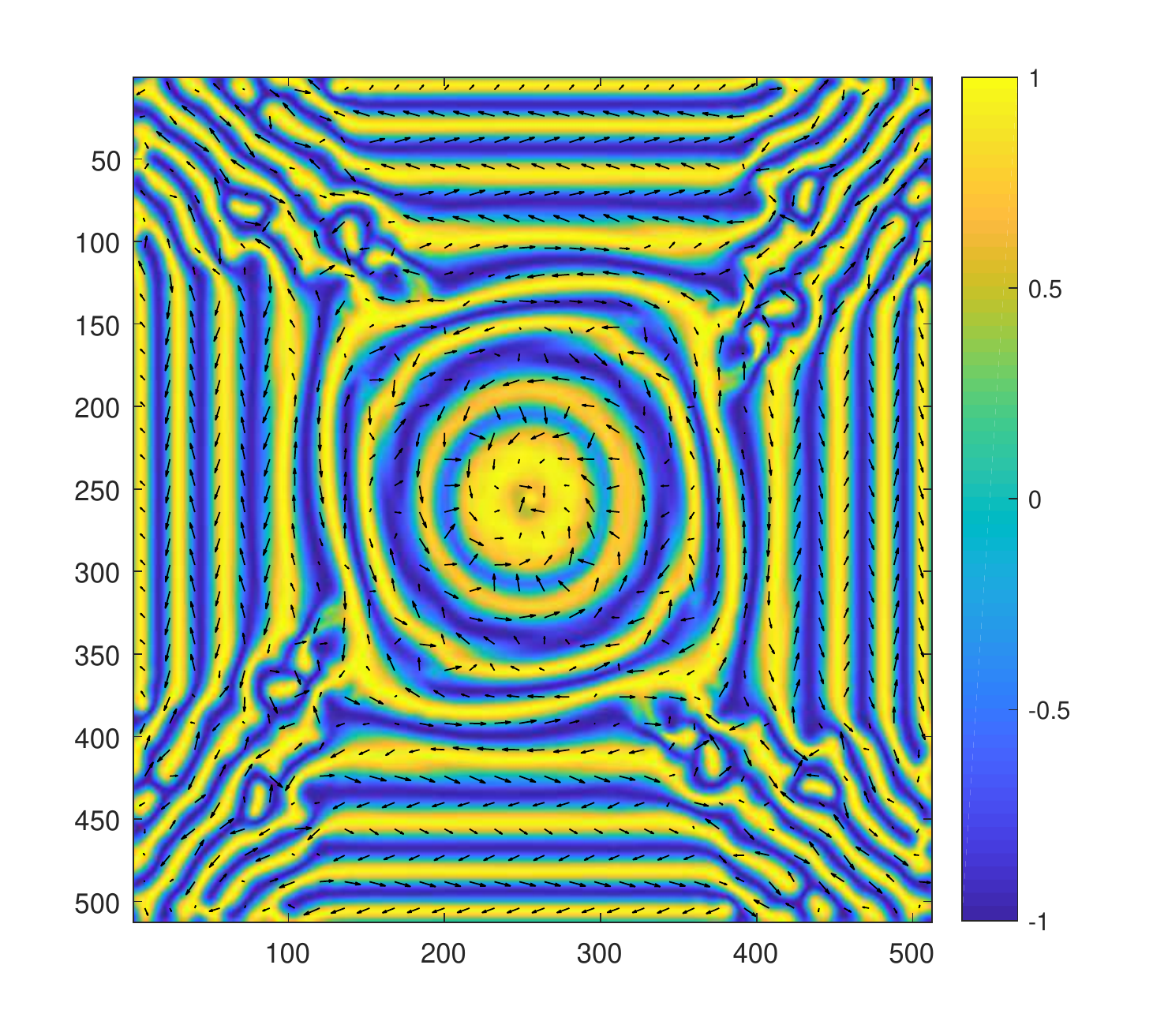}}\vfill
\subfloat[$\tau=800$]{\includegraphics[width=0.33\textwidth]{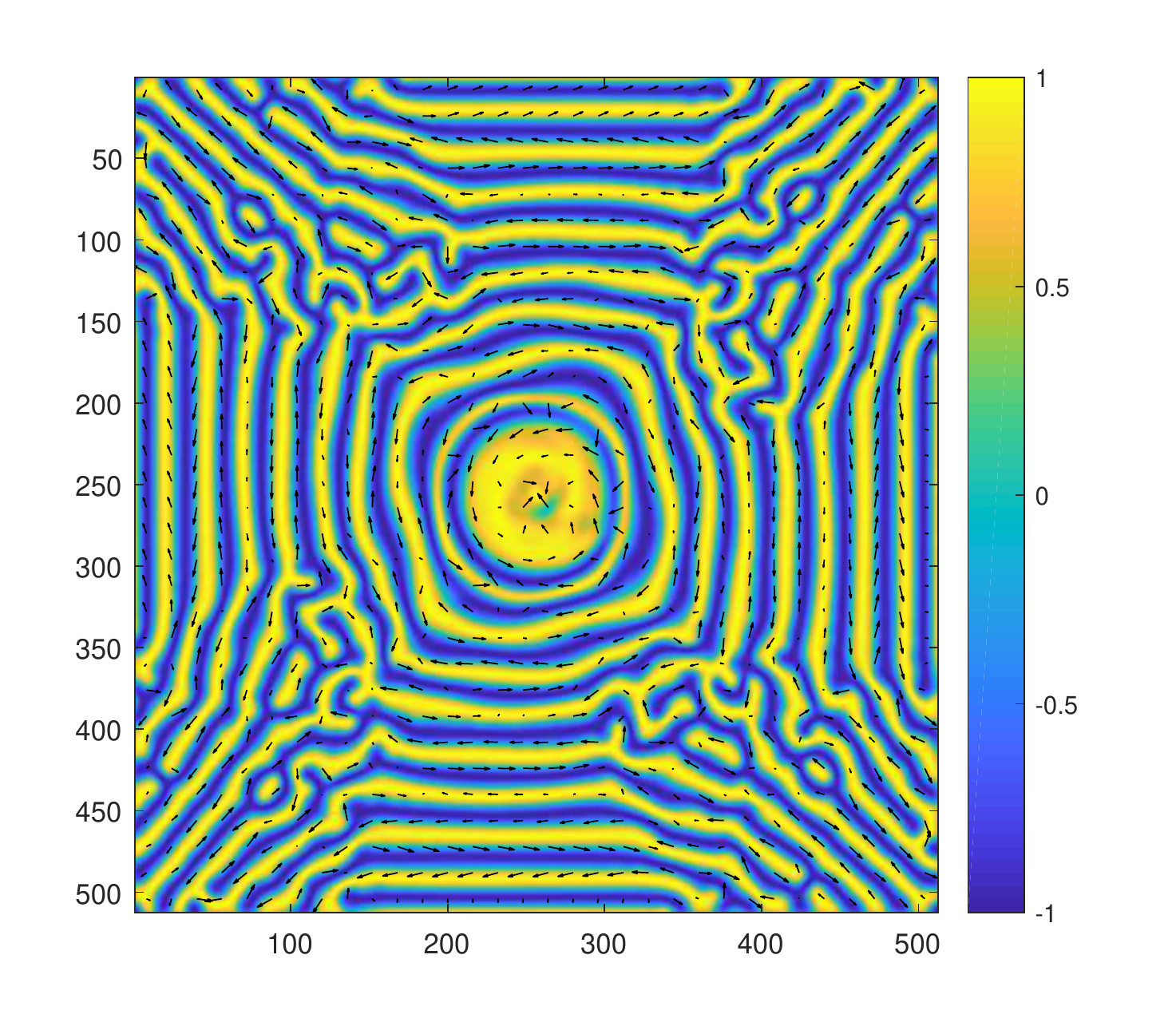}}\hfill
\subfloat[$\tau=1000$]{\includegraphics[width=0.33\textwidth]{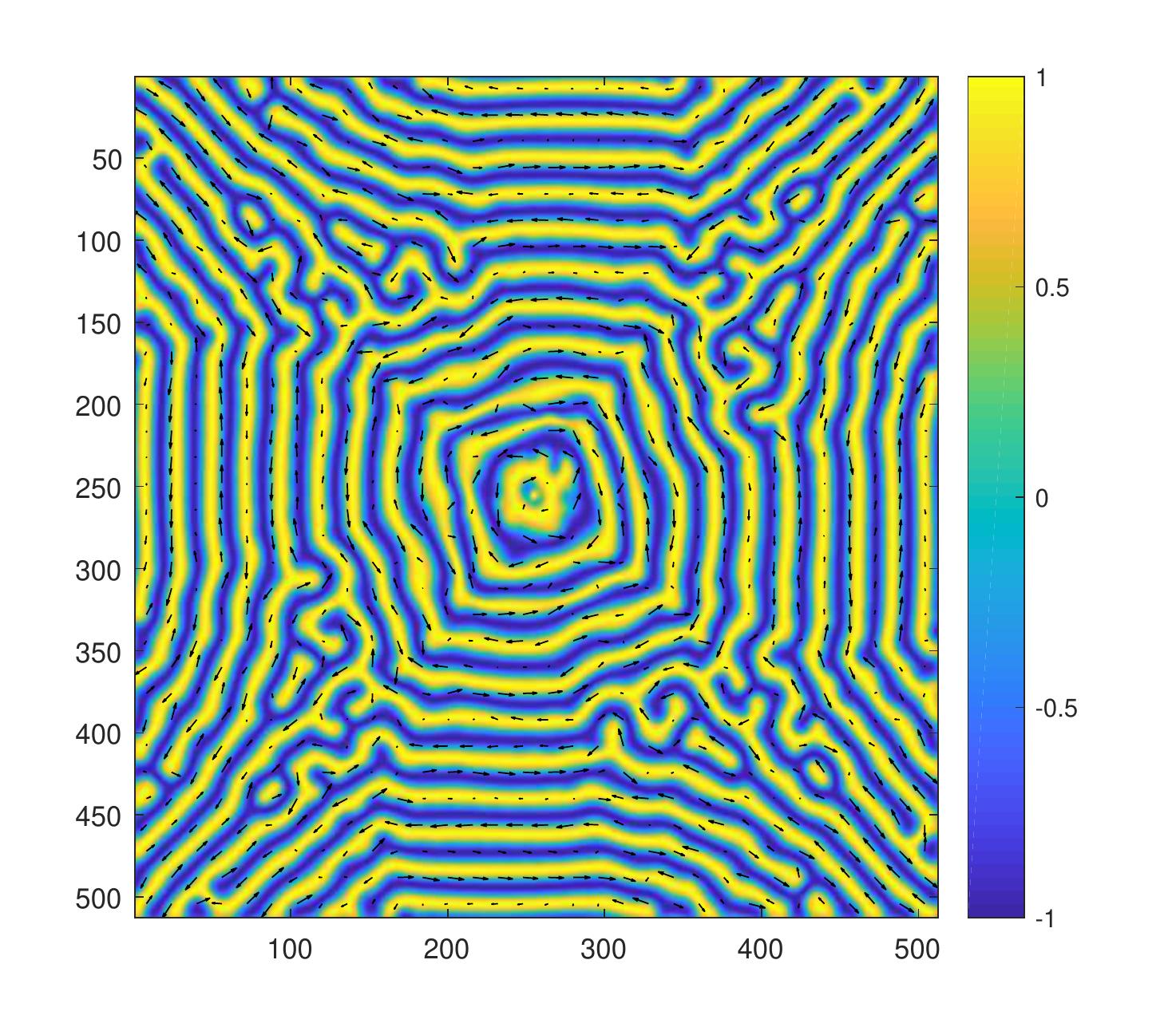}}\hfill
\subfloat[$\tau=10000$]{\includegraphics[width=0.33\textwidth]{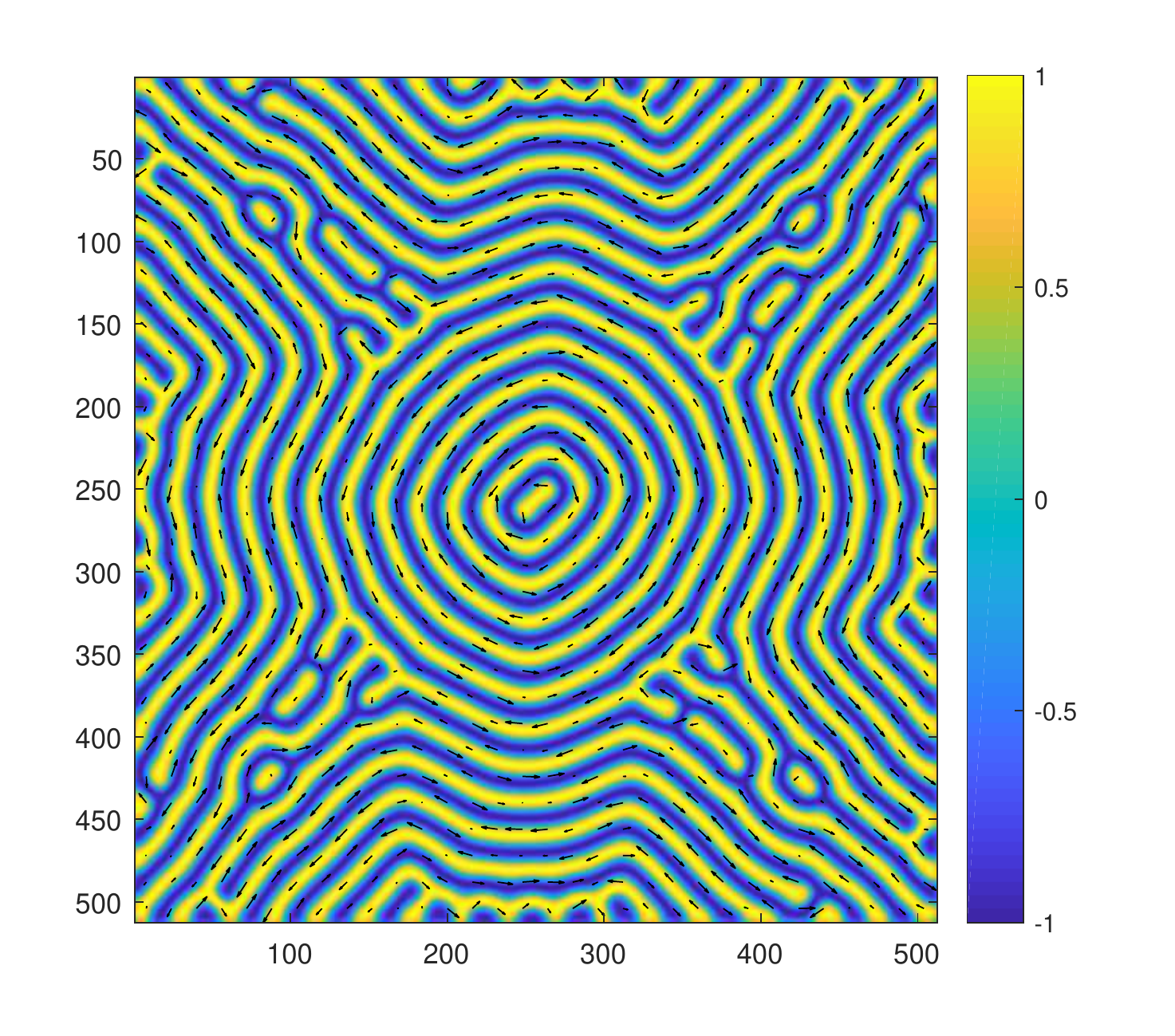}}\vfill
\caption{Simulation for $D/J=0.3,J_1=-0.5,J_2=0.012$. Because of the inhomogeneous $D$, a few domain walls are generated as rings around the pinning and keep expanding. The center of the smallest ring is a skyrmion. The largest ring keeps expanding until it meets the domain walls generated from the boundary. The smaller rings keep expanding until they are constrained by the larger rings. Finally, the center of the smallest ring becomes a skyrmion.}
\label{fig:satD03}
\end{figure}

For $D/J=0.1$, $0.2$, $0.3$, $0.5$, with the same values of other parameters as for $D/J=0.08$, stable skyrmions cannot be generated, because the skyrmion generated at the center  keeps expanding and becomes very large before it meets the domain walls generated from the boundary, so its border  splits into stripes. However, if we use $J_1<0$, domain walls are generated as several rings around the pinning and expand, as shown in Fig.~\ref{fig:satD03} for $D/J=0.3$. Then if the pinning radius is close to  the stripe widths, the skyrmions can be generated at the pinning center inside the smallest ring.

\begin{figure}
\subfloat[$D/J=0.1,J_1=-0.5,J_2=0.00156,\tau=16000$]{\includegraphics[width=0.33\textwidth]{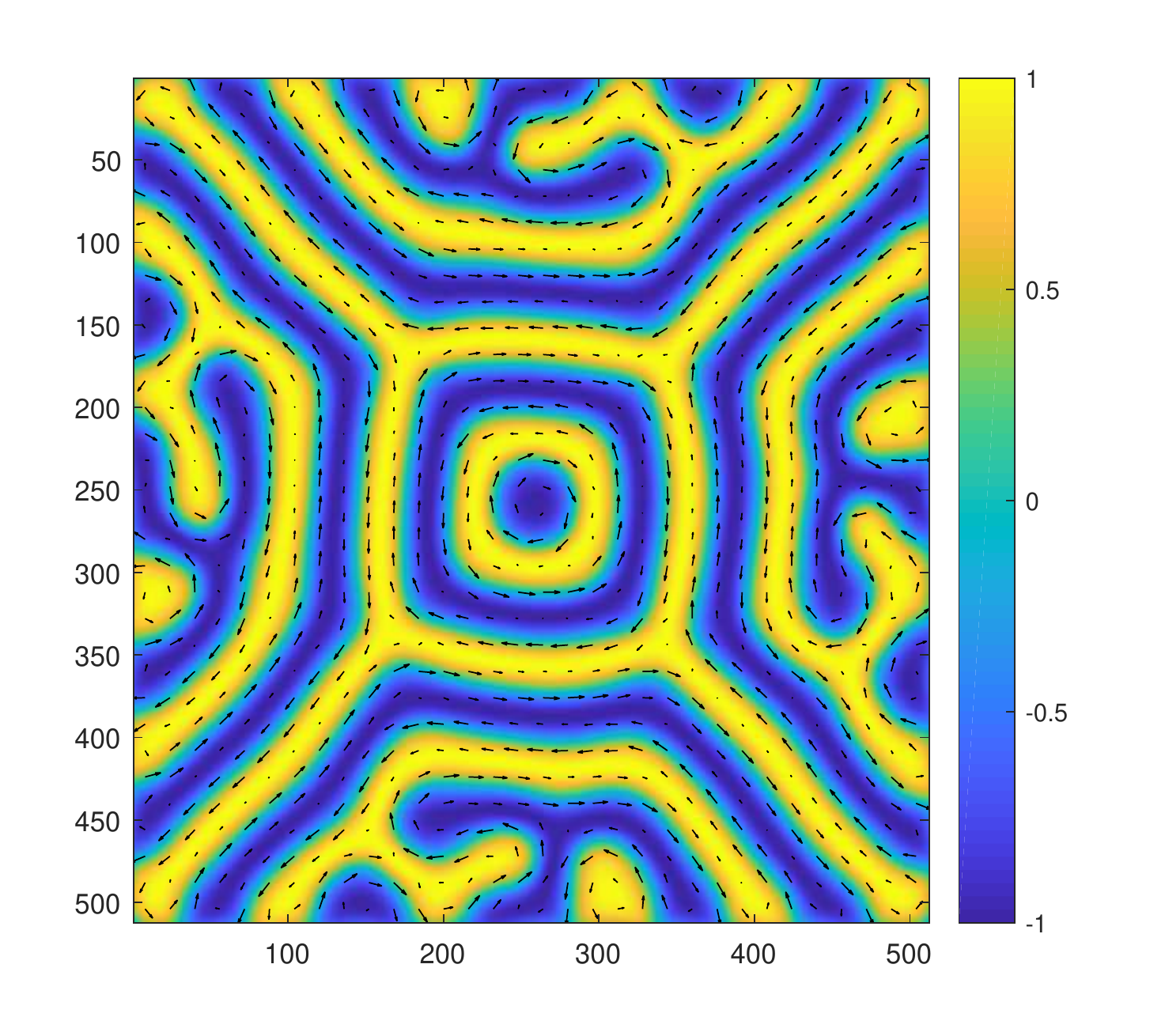}}\hfill
\subfloat[$D/J=0.2,J_1=-0.5,J_2=0.00625,\tau=8000$]{\includegraphics[width=0.33\textwidth]{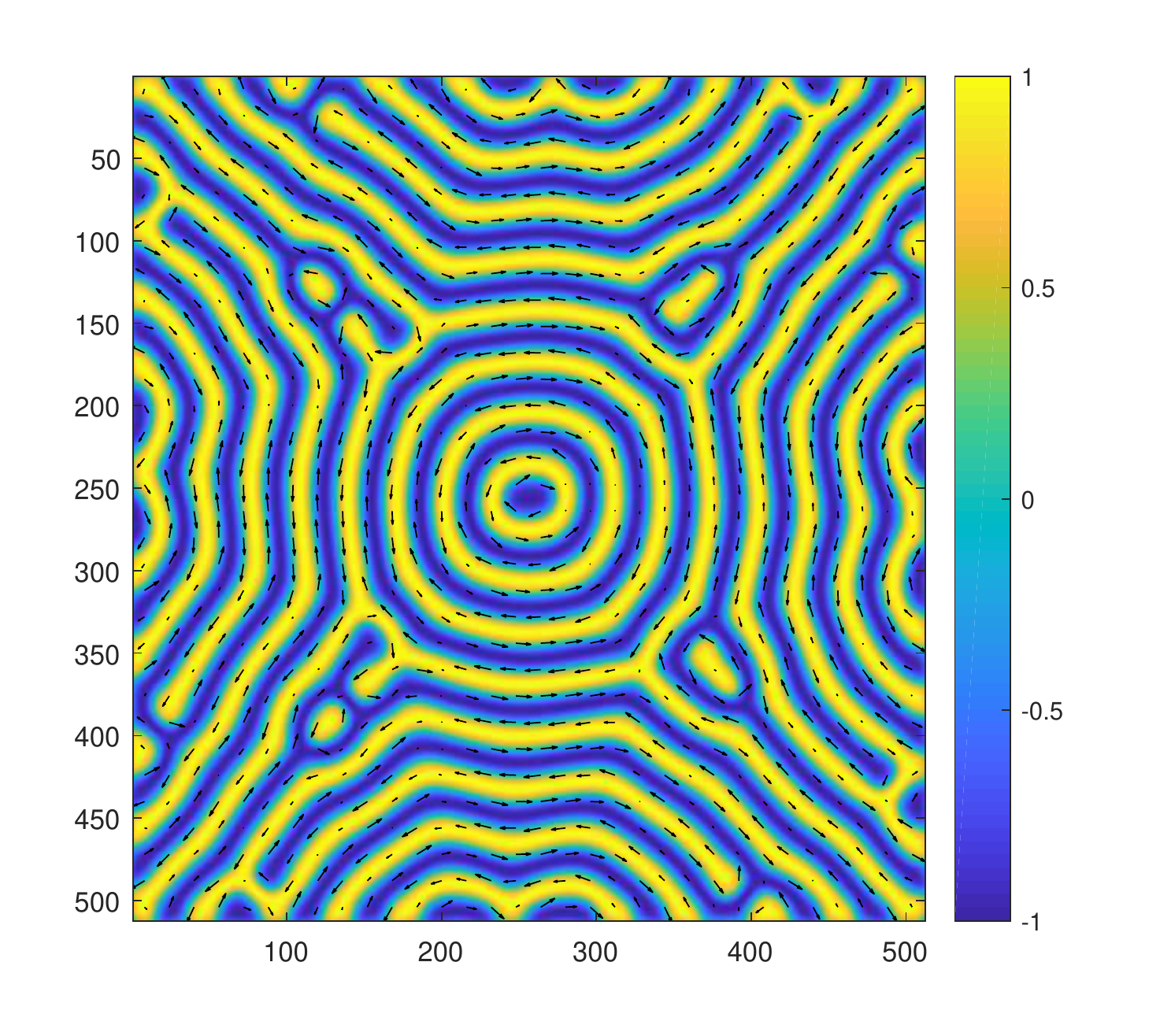}}\hfill
\subfloat[$D/J=0.5,J_1=-0.85,J_2=0.045,\tau=10000$]{\includegraphics[width=0.33\textwidth]{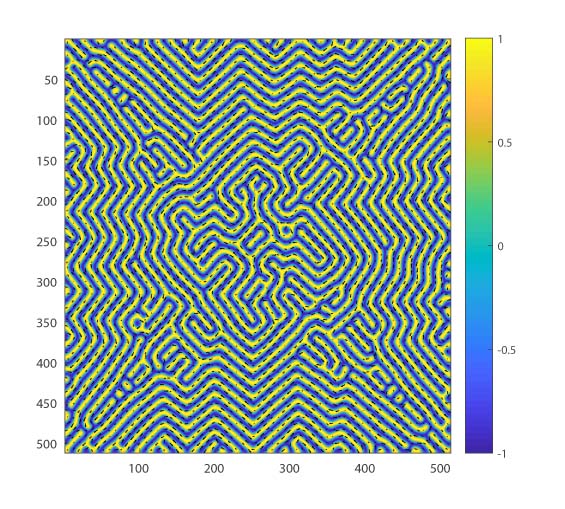}}\vfill
\caption{Simulation for $J_1<0$. For the case $D/J=0.5$, there is a short stripe cut off from all the others. It is the center of the smallest ring similar to  Fig.\ref{fig:satD03}, which can be considered as a stretched skyrmion.}
\label{fig:sats}
\end{figure}

We run the simulation for $D/J=0.1,0.2,0.3,0.5$ with $\alpha=0.1,\Delta t=0.002$ and $J_2=0.00156,0.00625,0.012,0.045\sim 1/D^2$. For $J_1<0$, $J_1$ is related to the gap between the rings.  For $D/J=0.1,0.2,0.3$, we use $J_1=-0.5$.  For $D/J=0.5$, the gap is too large compared with the stripe widths, so we use $J_1=-0.85$ to make gap narrower. The result for $D/J=0.3$ is shown in Fig.~\ref{fig:satD03} and the results for $D/J=0.1,0.2,0.5$ are shown in Fig.~\ref{fig:sats}.

\begin{figure}
\subfloat[$\tau=2000$]{\includegraphics[width=0.33\textwidth]{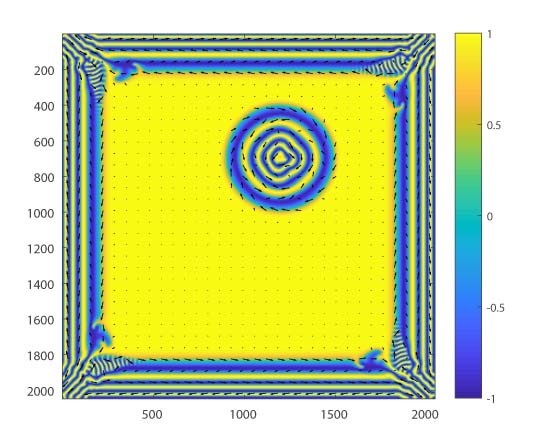}}\hfill
\subfloat[$\tau=3000$]{\includegraphics[width=0.33\textwidth]{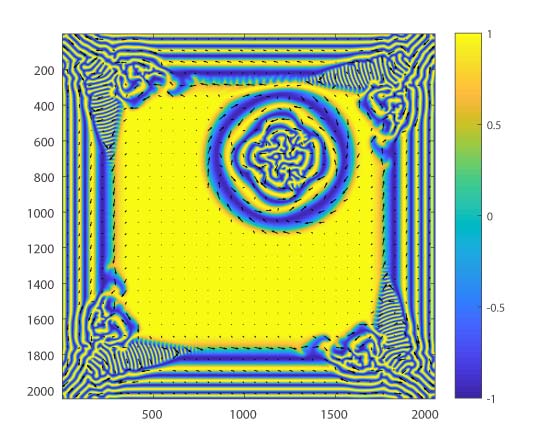}}\hfill
\subfloat[$\tau=4000$]{\includegraphics[width=0.33\textwidth]{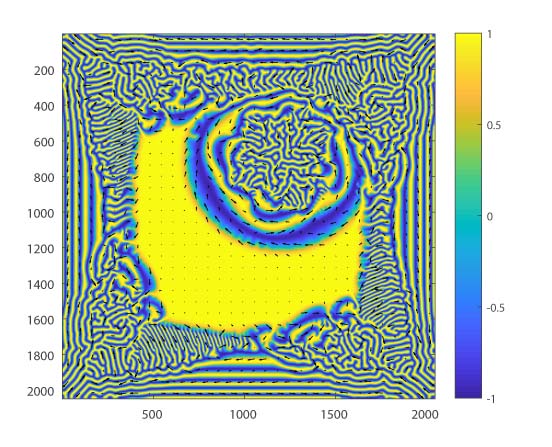}}\vfill
\subfloat[$\tau=5000$]{\includegraphics[width=0.33\textwidth]{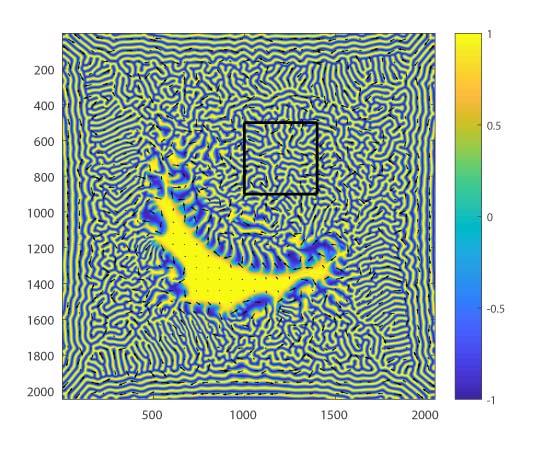}}\hfill
\subfloat[$\tau=15000$]{\includegraphics[width=0.33\textwidth]{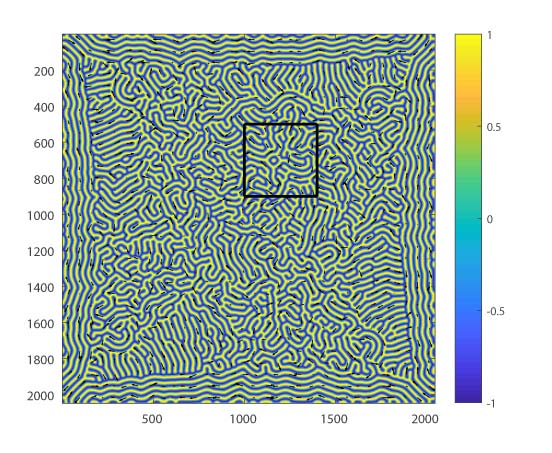}}\hfill
\subfloat[$\tau=15000$]{\includegraphics[width=0.33\textwidth]{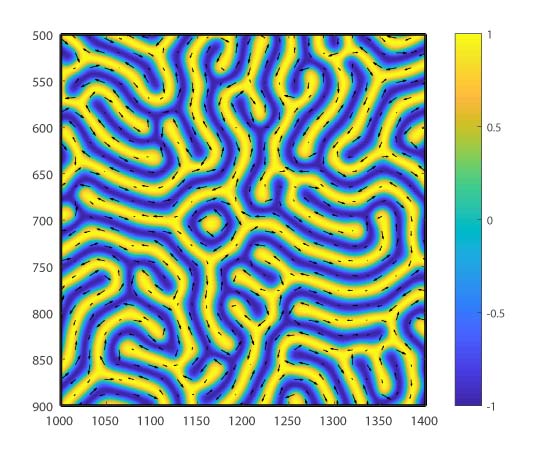}}\vfill
\caption{Simulation on a $2048\times 2048$ lattice with $D/J=0.2, \alpha = 0.2$, $\Delta t =0.01$, $J_1=-0.8$, $J_2=0.005$, and the center of the pinning placed at $(1200,700)$. First, the rings appear and expand. At about $\tau = 3000$, the rings meet the domain walls generated from the boundary. The center of the rings becomes stable at about $\tau=5000$. Finally, the central region of the smallest ring survives as  a skyrmion, however, its position  deviates from the   the pinning center.}
\label{fig:satLarge}
\end{figure}

Comparing the final states in Sec.\ref{sec:4.3} with the initial states in Sec.\ref{sec:4.2}, we can conclude that, with boundary condition alone, the skyrmions are not generated at the center. Therefore, the effect of the pinning is essential.  We also run a simulation to testify how important the boundary condition is. In real experiments, the material could be much larger, and the pinning could be located away from the center of the material. We run a simulation on a $2048\times 2048$ lattice with the pinning located at $(1200,700)$ to investigate  the generation of the skyrmion in this situation. We choose $D/J=0.2$, and follow the rescaling method in Refs.~\cite{Tchoe,pin}. If for the real material, $\lambda \approx 60 \;{\rm nm}$, $a\approx 4\;{\AA}$, the rescaling factor is  $r\approx 3.38$, therefore, a $2048 \times 2048$ lattice corresponds to about $2.77\;{\rm \mu m}\times 2.77\; {\rm \mu m}$.  For $J=3 \;{\rm meV}$, the dimensionless time $\tau=15000$ corresponds to $37.6\; {\rm n s}$. We run the simulation with $\alpha = 0.2$, $\Delta t =0.01$, $J_1=-0.8$ and $J_2=0.005$. The result is shown in Fig.\ref{fig:satLarge}. We find that, a skyrmion is generated  on the left of the  pinning  center, and the distance between the   skyrmion center and  the pinning center  is about $50$,  corresponding to  $67.5\;{\rm nm}$ if we use the above realistic parameter values.

\subsection{\label{sec:4.4}Bound states}

\begin{figure}
\subfloat[$\tau=100000$]{\includegraphics[width=0.33\textwidth]{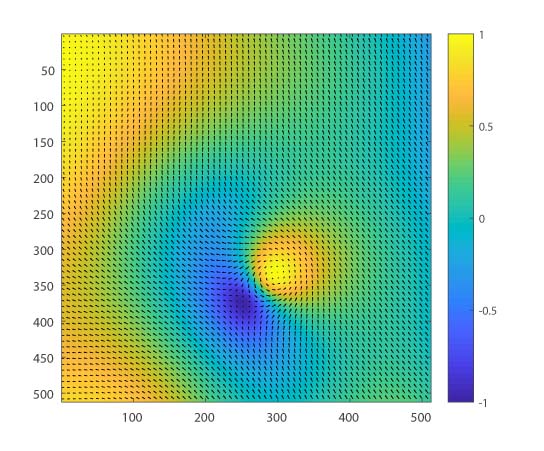}}\hfill
\subfloat[$\tau=120000$]{\includegraphics[width=0.33\textwidth]{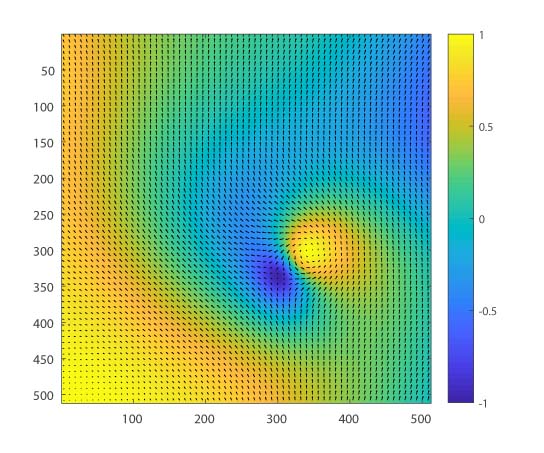}}\hfill
\subfloat[$\tau=140000$]{\includegraphics[width=0.33\textwidth]{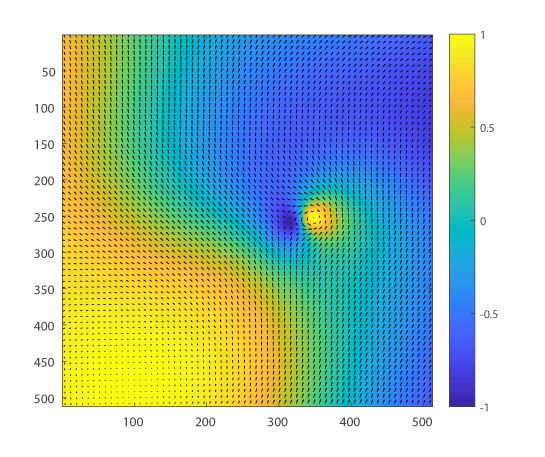}}\vfill
\subfloat[$\tau=160000$]{\includegraphics[width=0.33\textwidth]{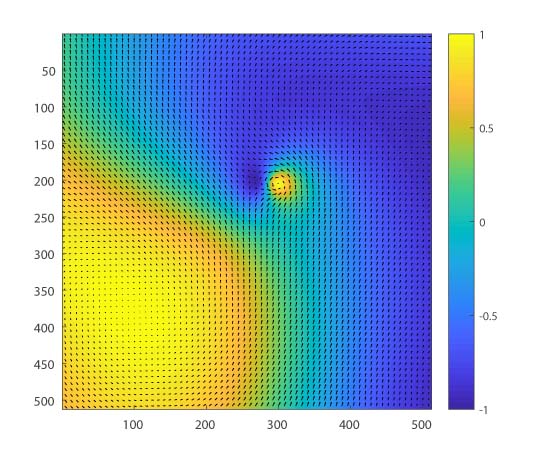}}\hfill
\subfloat[$\tau=180000$]{\includegraphics[width=0.33\textwidth]{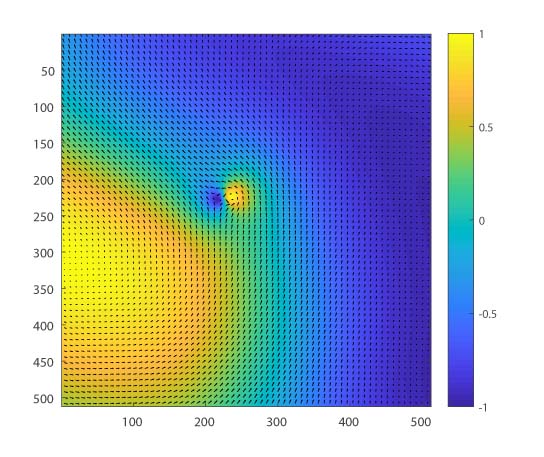}}\hfill
\subfloat[$\tau=200000$]{\includegraphics[width=0.33\textwidth]{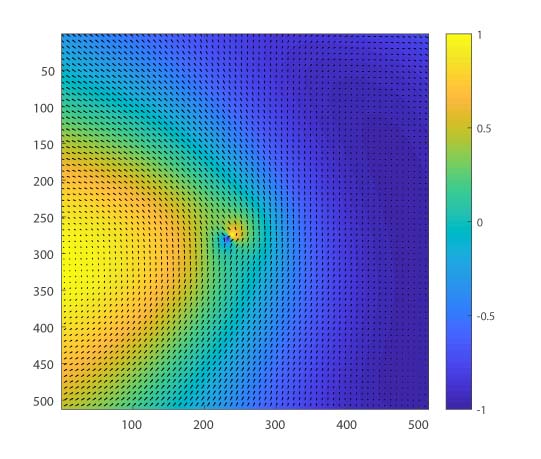}}\vfill
\caption{Simulation of skyrmion generation, with parameter values  $J_1=-0.5$, $J_2=0.0001$, $D=0.005$, $\alpha=0.2$ and $\Delta t=0.01$.  A skyrmion and an anti-skyrmion with $g=\pm 1$ and $m=\pm 1$ respectively rotate around each other and move to the pinning center, with the distance shrinking till annihilation. }
\label{fig:special1}
\end{figure}

\begin{figure}
\subfloat[$\tau=100000$]{\includegraphics[width=0.33\textwidth]{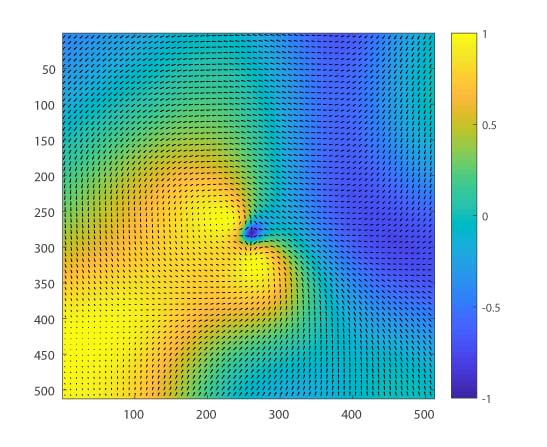}}\hfill
\subfloat[$\tau=120000$]{\includegraphics[width=0.33\textwidth]{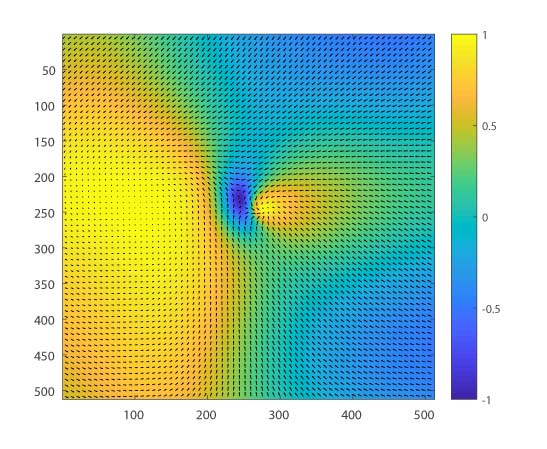}}\hfill
\subfloat[$\tau=140000$]{\includegraphics[width=0.33\textwidth]{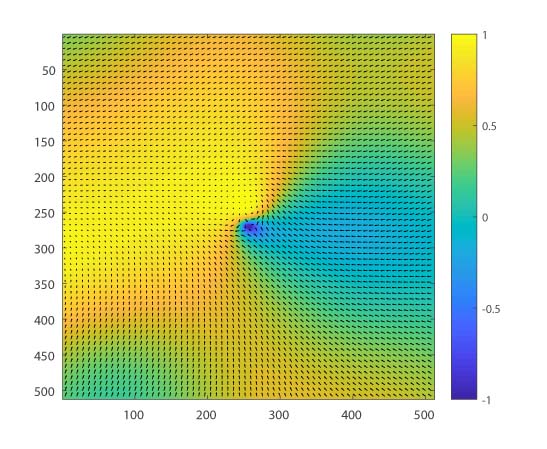}}\vfill
\subfloat[$\tau=160000$]{\includegraphics[width=0.33\textwidth]{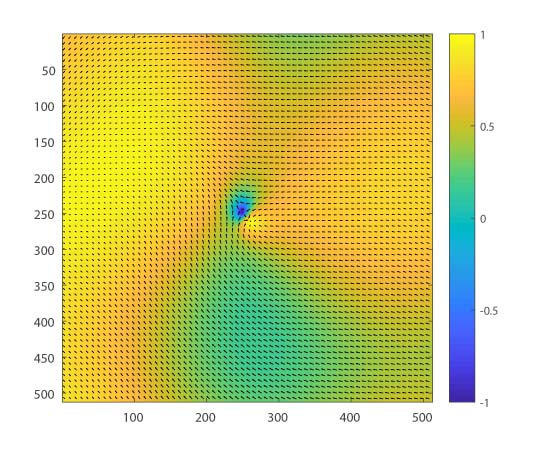}}\hfill
\subfloat[$\tau=180000$]{\includegraphics[width=0.33\textwidth]{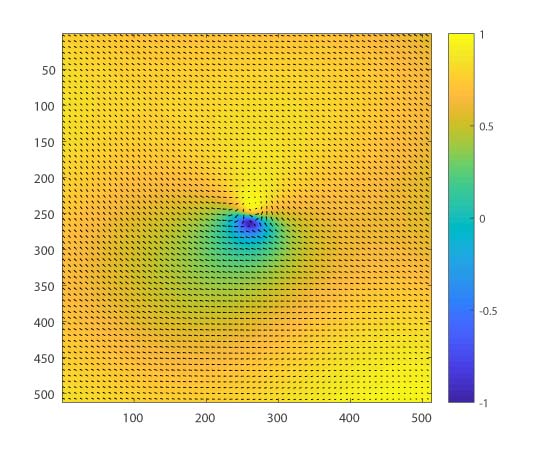}}\hfill
\subfloat[$\tau=200000$]{\includegraphics[width=0.33\textwidth]{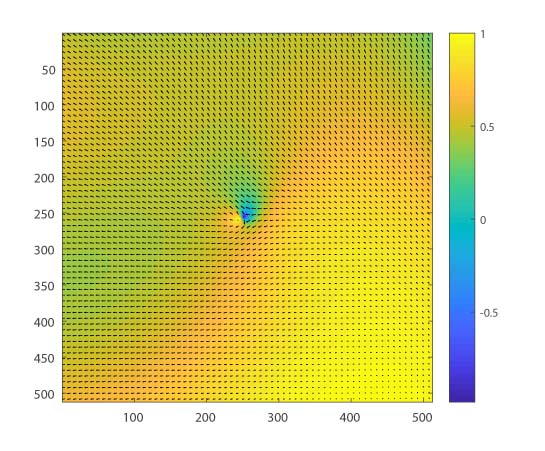}}\vfill
\caption{Simulation of skyrmion generation, with parameter values $J_1=-0.9$, $J_2=0.0001$, $D=0.003$, $\alpha=0.1$ and $\Delta t=0.01$. A skyrmion and an anti-skyrmion, with $g=\pm 1$ and $m=\pm 1$  respectively, are generated at the pinning center. Each skyrmion self-rotates. The simulation stopped at $\tau=200000$, however. As the distance  shrinks, the pair is expected  to annihilate.}
\label{fig:special2}
\end{figure}

In  the simulation starting from randomized initial state, we also observe an interesting phenomenon when $D$ is constant and very small while $J_1<0$ (Figs.~\ref{fig:special1} and ~\ref{fig:special2}). It  does not show up when  $J_1>0$ or    $J_1=0$ (no pinning), so it  is a special case in presence of pinning with   $J_1<0$. When $D$ is very small, a skyrmion and an anti-skyrmion with $g=\pm 1$ and $m=\pm 1$ are generated and move to the pinning center. They keep rotating around each other with the distance shrinking till annihilation. Previously, it was found that the interaction between two skyrmions on two layers with opposite skyrmion charge can form a bound state~\cite{bilayer}.

On a single layer, in contrast, if the skyrmions are generated in an external magnetic field, the sign of $g$ is determined by the external magnetic field, and whether a skyrmions or an anti-skyrmions is generated depends on $D$, then the  bound states of skyrmions and anti-skyrmions are usually difficult to realize.  The situation we consider may provide a novel avenue  to study such bound states in a single layer.


\section{\label{sec:5} Summary  }

In this paper, we propose a novel mechanism to generate magnetic skyrmions without the need of an external magnetic field or  magnetic anisotropy. We find that skyrmions can be generated through the pinning effect only, i.e., with   magnetic exchange strength $J$ inhomogeneous, or with $J$ and DM interaction $D$ both inhomogeneous. Our lattice simulation has verified this idea. In the simulation, we study the properties of the skyrmions generated under various parameter values. We find that the radius of the skyrmion increases when $D$ decrease. We also find that all the skyrmions generated have $m=1$ and $\tilde{D}>0$, while the sign of $g$ depends on the initial state. For $J_1<0$, we also find the generation of a pair of skyrmions with opposite $g$ and $m$ at the pinning centers.

That the skyrmions  can be generated  by using the pinning effect  only is  useful  for practical applications of the magnetic skyrmions. Through the engineering of  pinning in the designated site, we can generate a skyrmion on this site. It is hoped that  experiments and applications be made by using this method.

This work is supported by National Natural Science Foundation of China (Grant No. 11374060 and No. 11574054).


\clearpage
\begin{thebibliography}{99}
\bibliographystyle{unsrt}

\bibitem{nagaosa}
N. Nagaosa and Y. Tokura, Nat. Nanotechnol. \textbf{8}, 899 - 911 (2013).

\bibitem{Fert}
A. Fert, N. Reyren and V. Cros, Nat. Rev. Mater. \textbf{2}, 17031 (2017).

\bibitem{Muhlbauer}
S. M\"{u}hlbauer et. al. Science \textbf{323}, 915 - 919 (2009).

\bibitem{Yu1}
X. Z. Yu et. al. Nature \textbf{465}, 901 - 904 (2010).

\bibitem{2DSkyrmion2}
S. Heinze et. al. Nature Phys. \textbf{7}, 713 - 718 (2011);

C. Pfleiderer, Nature Phys. \textbf{7}, 673 - 674 (2011).

\bibitem{DMI}
I. Dzyaloshinskii, J. Phys. Chem. Solids \textbf{4}, 241 - 255 (1958);

T. Moriya, Phys. Rev. \textbf{120}, 91 - 98 (1960).

\bibitem{ultralow}
F. Jonietz et. al. Science \textbf{330}, 1648 - 1651 (2010);

X. Z. Yu et. al. Nature Commun. \textbf{3}, 988 (2012).

\bibitem{Romming}
N. Romming et. al. Science \textbf{341}, 636 - 639 (2013).

\bibitem{bubble}
W. Jiang et. al. Science \textbf{349}, 283 - 286 (2015).

\bibitem{thetarho}
W. Jiang et. al. Physics Reports, {\bf 704}, 1 - 49 (2017), arXiv:1706.08295.

\bibitem{Tchoe} 
Y. Tchoe and J. H. Han, Phys. Rev. B. {\bf 85}, 174416 (2012), arXiv:1203.0638.

\bibitem{anisotropy}
A. Bogdanov and A. Hubert, J. Magn. Magn. Mater. {\bf 195} 182 (1999).

\bibitem{bilayer1}
X. Zhang et. al. Nat. Commun. \textbf{7}, 10293 (2016), arXiv:1504.02252.

\bibitem{dccurrent}
K. Everschor-Sitte et. al. New J. Phys. {\bf 19}, 092001, (2017), arXiv:1610.08313.

\bibitem{anisotropyenergy}
P. Eames and E. Dan Dahlberg, J. Appl. Phys. {\bf 91}, 7986 (2002);

C. Moutafis et. al. Phys. Rev. B \textbf{74}, 214406 (2006).

S. Finizio et. al. Phys. Rev. B \textbf{98}, 104415 (2018).

\bibitem{pin} 
Y.-H. Liu and Y.-Q. Li, J. Phys. Condens. Matter {\bf 25} 076005 (2013).

\bibitem{pin2}
S. Z. Lin et. al. Phys. Rev. B \textbf{87}, 214419 (2013).

\bibitem{FreeEnergy}
J. H. Han et. al. Phys. Rev. B {\bf 82}, 094429 (2010).

\bibitem{antiskyrmion}
S. Huang et. al. Phys. Rev. B {\bf 96} 144412 (2017); 

\bibitem{arctan} 
R. Nepal, U. G\"{u}ng\"{o}rd\"{u} and A. A. Kovalev, Appl. Phys. Lett. {\bf 112}, 112404 (2018), arXiv:1711.03041.

\bibitem{unitJ}
M. Mochizuki, Phys. Rev. Lett. {\bf 108}, 017601 (2012), arXiv:1111.5667. 

\bibitem{alpha3}
C. Sch\"{u}tte et. al. Phys. Rev. B {\bf 90} 174434 (2014). 

\bibitem{bilayer}
W. Koshibae and N. Nagaosa, Sci. Rep. {\bf 7} 42645 (2017). 

\bibitem{LLG}
G. Tatara, H. Kohno and J. Shibata, Phys. Rep. {\bf 468}, 213 - 301 (2008), arXiv:0807.2894;

\bibitem{LLG2}
J. Zang et. al. Phys. Rev. Lett. {\bf 107} 136804 (2011). 

\bibitem{discreteH}
J. Iwasaki, M. Mochizuki, and N. Nagaosa, Nat. Nanotechnol. \textbf{8}, 742 - 747, (2013),  arXiv:1310.1655; 

J. Iwasaki, M. Mochizuki, and N. Nagaosa, Nature Commun. \textbf{4}, 1463, (2013); 

\bibitem{alpha1}
J. Sampaio et. al. Nat. Nanotechnol. \textbf{8}, 839, (2013); 

K. Litzius et. al. Nat. Phys. \textbf{13}, 170 - 175 (2016); 

W. Jiang et. al. Nat. Phys. \textbf{13}, 162 - 169 (2016), arXiv:1603.07393; 

\bibitem{alpha2}
R. Tomasello et. al. J. Phys. D: Appl. Phys. {\bf 50} 325302 (2017); 

S. H. Yang et. al. Nat. Nanotechnol. \textbf{10}, 221, (2015); 

J. Barker and O. A. Tretiakov, Phys. Rev. Lett. {\bf 116} 147203 (2016). 

\bibitem{dvalue}
L. Kong and J. Zang, Phys. Rev. Lett. {\bf 111} 067203 (2013); 

j. Iwasaki, W. Koshibae and N. Nagaosa, Nano. Lett. {\bf 14}, 4432-7 (2014). 


\end{thebibliography}
\end{document}